\documentclass[letterpaper,11pt]{article}

\usepackage{jheppub}
\usepackage{subfig}

\def\Fig#1{fig.~{\ref{#1}}}

\DeclareRobustCommand{\Sec}[1]{sec.~\ref{#1}}

\def\cE{\mathcal{E}}

\newcommand{\ga}{\gamma}

\newcommand{\bea}{\begin{eqnarray}}
\newcommand{\eea}{\end{eqnarray}}

\def\cE{\mathcal{E}}

\newcommand{\cQ}{\mathcal{Q}}

\def \as {\relax\ifmmode\alpha_s\else{$\alpha_s${ }}\fi}

\newcommand{\df}{\mathrm{d}}
\newcommand{\nn}{\nonumber}

\allowdisplaybreaks[2]

\definecolor{darkgreen}{rgb}{0.13,0.55,0.13}

\begin{document}
\preprint{MIT-CTP-5591}

\title{Joint Track Functions: Expanding the Space of Calculable Correlations at Colliders}

\author[1]{Kyle Lee}
\author[2]{and Ian Moult}

\affiliation[1]{Center for Theoretical Physics, Massachusetts Institute of Technology, Cambridge, MA 02139}
\affiliation[2]{Department of Physics, Yale University, New Haven, CT 06511, USA\vspace{0.5ex}}

\abstract{The theoretical description of observables at collider experiments relies on factorization theorems separating perturbative dynamics from universal non-perturbative matrix elements. 
Despite significant recent progress in extending these factorization theorems to increasingly differential jet substructure observables, the focus has been primarily on infrared safe observables sensitive only to correlations in the energy of final state hadrons.
However, significant information about the dynamics of the underlying collision is encoded in how energy is correlated between hadrons of different quantum numbers.
In this paper we extend the class of calculable correlations by deriving factorization theorems for a broad class of correlations, $\langle \mathcal{E}_{R_1}(n_1) \cdots \mathcal{E}_{R_k}(n_k) \rangle$, between the energy flux carried by hadrons specified by quantum numbers, $R_1, \cdots, R_k$.
We show that these factorization theorems involve moments of a new class of universal non-perturbative functions, the ``joint track functions", which extend the track function formalism to describe the fraction of energy carried by hadrons of multiple quantum numbers arising from the fragmentation of quarks or gluons. 
We study the general properties of these functions, and then apply this to the specific case of joint track functions for positive and negative electromagnetic charges. We extract these  from parton shower Monte Carlo programs and use them to calculate correlations in electromagnetically charged energy flux. We additionally propose and study a C-odd $\mathcal{E}_{\mathcal{Q}}$ detector, which results in a qualitatively distinct scaling behavior compared to the standard energy correlators.
Our formalism significantly extends the class of observables that can be computed at hadron colliders, with a wide range of applications from particle to nuclear physics.
}

\maketitle

%%%%%%%%%%%%%%%%%%%%%%
\section{Introduction}\label{sec:intro}
%%%%%%%%%%%%%%%%%%%%%%

Collider physics experiments present a theoretical challenge due the hierarchy of scales present between the underlying hard scattering scale, $Q$, at which the microscopic interactions occur, and the infrared scale, $\Lambda$, at which measurements are performed. This is particularly true when the underlying theory is not scale invariant, such as in quantum chromodynamics (QCD), which exhibits a complete rearrangment of the degrees of freedom, from quarks and gluons to hadrons, between these two scales. Our ability to theoretically describe collider experiments relies on factorization theorems \cite{Collins:1989gx,Collins:1988ig,Collins:1985ue}, which rigorously separate the perturbatively calculable physics of the microscopic interactions of quarks and gluons from universal non-perturbative matrix elements describing the hadronization process.

One of the primary advances of the Large Hadron Collider (LHC) has been the advent of jet substructure \cite{Larkoski:2017jix,Kogler:2018hem}, which uses correlations in the asymptotic energy flux to learn about the underlying microscopic dynamics of the theory.  Its theoretical description has required the development of increasingly sophisticated factorization theorems, see e.g. \cite{Fleming:2007qr,Fleming:2007xt,Becher:2008cf,Abbate:2010xh,Larkoski:2015kga,Frye:2016aiz,Larkoski:2017iuy,Hoang:2017kmk,Larkoski:2017cqq}. At a technical level, this has been enabled primarily by the development of powerful effective field theory techniques for proving factorization \cite{Bauer:2000ew, Bauer:2000yr, Bauer:2001ct, Bauer:2001yt,Rothstein:2016bsq}. Nonetheless, it has, with several notable exceptions, e.g. \cite{Krohn:2012fg,Waalewijn:2012sv}, been restricted to observables that are only sensitive to the energy of final state hadrons. Such observables are infrared and collinear safe \cite{Kinoshita:1962ur,Lee:1964is}, and hence exhibit the simplest form of factorization theorem.

Theoretically favorable measurements of jet substructure are formulated in terms of the energy flux operator \cite{Sveshnikov:1995vi,Tkachov:1995kk,Korchemsky:1999kt,Bauer:2008dt,Hofman:2008ar,Belitsky:2013xxa,Belitsky:2013bja,Kravchuk:2018htv},
\begin{align}
    \cE(\vec n_1) = 
    \lim_{r\rightarrow \infty} \int \mathrm{d}t \,r^2 n_1^i \,
    T_{0i}(t,r\vec{n}_1)\,,
    \label{eq:def}
\end{align}
which can be viewed as a theoretical idealization of a calorimeter cell measuring the energy flux of \emph{all} hadrons. Multi-point correlation functions of these operators are infrared and collinear safe $\langle \cE(n_1) \cdots  \cE(n_k) \rangle$ \cite{Basham:1979gh,Basham:1978zq,Basham:1978bw,Basham:1977iq}. These observables have recently seen widespread phenomenological application to the study of jet substucture \cite{Dixon:2019uzg,Chen:2020vvp,Komiske:2022enw,Holguin:2022epo,Craft:2022kdo,Devereaux:2023vjz,Chen:2022swd,Chen:2019bpb,Andres:2022ovj,Lee:2022ige,Andres:2023xwr,Liu:2022wop,Liu:2023aqb,Cao:2023rga}. They exhibit rigorous factorization theorems in $e^+e^-$ collisions \cite{Dixon:2019uzg}, and at hadron colliders \cite{Lee:2022ige}. Reformulating jet substructure in terms of correlation functions has led to a much closer connection between experimental measurements, and the theoretical description of the underlying quantum field theory in terms of correlation functions.

\begin{figure}
\begin{center}
\includegraphics[scale=0.3]{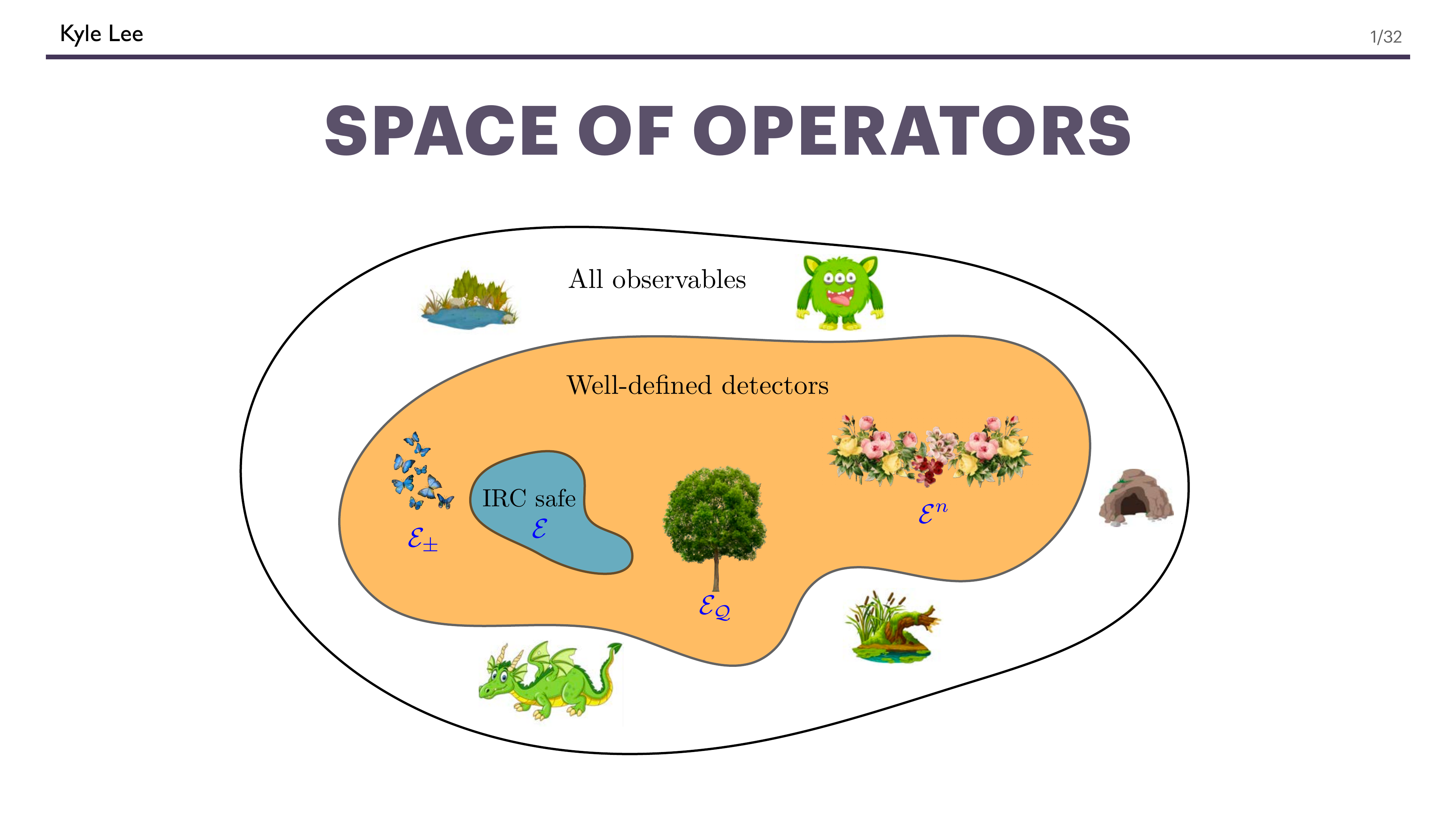} 
\end{center}
\caption{An illustration presenting the space of field theoretically well-defined detectors. By studying observables with these detectors, we can simultaneously expand the landscape of possible observables, enhance the sensitivity to nontrivial hadronization effects, and retain perturbative control.}
\label{fig:detectorspace}
\end{figure} 

However, the energy correlators are only a small class of the correlations in the energy flux of final state hadrons that can be measured experimentally.  More detailed properties of the energy flux can be studied by defining operators that measure the charge flux on hadrons with a given property. For example, one can define $\cE_{\pm}(n)$ to be the operator that measures the energy flux on positively/negatively charged hadrons. A particularly interesting set of quantities formulated in terms of these operators are correlation functions of the form $\langle \cE_+(n_1) \cE_-(n_2) \rangle$. These observables can provide insight into how quantum numbers are correlated during the hadronization process, see e.g. \cite{Chien:2021yol}. However, the calculation of these observables, which are sensitive to the quantum numbers of the spectrum of hadronic states, is more challenging theoretically. In particular, factorization theorems describing these observables must incorporate new non-perturbative functions describing correlations in the hadronization process. The exploration of these more general factorization theorems is of intrinsic theoretical interest, extends the class of calculable observables, and explores how far the combination of factorization theorems and the renormalization group can be pushed.

Significant progress in deriving factorization theorems to expand the class of calculable correlations has recently been made using the track function formalism \cite{Chang:2013rca,Chang:2013iba,Chen:2022pdu,Chen:2022muj,Jaarsma:2022kdd,Li:2021zcf,Lee:2023xzv}. The track function formalism allows one to calculate correlation functions, $\langle \cE_R(n_1) \cdots \cE_R(n_k) \rangle$, on the energy flux carried by a hadrons specified by a particular quantum number, $R$, e.g. charged energy flux.  For example, for the case of electromagnetic charge, the track function formalism allows the calculation of $\langle \cE_+(n_1) \cE_+(n_2) \rangle$ and $\langle \cE_-(n_1) \cE_-(n_2) \rangle$ correlators. The track function formalism is particularly convenient for energy correlators, where it can be viewed as a matching between parton level detectors, and hadron level detectors. This is illustrated schematically in \Fig{fig:2stage}, where the parton level detectors are shown as measurements on the CMB, and the hadron level detectors are shown at the level of large scale structure, where the galaxies are analogous to hadrons. The track function formalism provides the non-perturbative matching coefficients between the detectors at the two stages. Similar to the CMB seeding the density perturbation for large-scale cosmic structures, partons, via the hadronization process, give birth to the intricate assembly of hadrons. However, the track function formalism does not allow the calculation of correlations between energy flux on different charges, e.g. $\langle \cE_+(n_1) \cE_-(n_2) \rangle$, as well as their generalizations to higher points. One would therefore like to extend this formalism, and introduce the necessary non-perturbative matrix elements that allow the calculation of these most general correlations. Only with such extension, we are able to probe nontrivial correlations between different quantum numbers in the hadronization process in Fig.~\ref{fig:jt_total}.

In this paper we introduce new non-perturbative matrix elements which allow the calculation of such correlations, and study their renormalization group structure, and their application to energy correlator observables. Our motivation is two-fold: First, we would like to push the boundary of what can be studied about the parton to hadron transition using perturbation theory and the renormalization group. Second, we would like to develop the necessary theoretical formalism to study energy correlators on different quantum numbers. We call the new non-perturbative objects ``joint track functions". These can be viewed as an extension of the track functions \cite{Chang:2013rca,Chang:2013iba} to measuring energy fractions on multiple class of hadrons. Indeed, their renormalization group evolution will be similar, and we can build on recent developments in the understanding of track functions  \cite{Li:2021zcf,Jaarsma:2022kdd,Chen:2022pdu,Chen:2022muj}. In some sense the extension from track functions to joint track functions  is analogous to the extension from fragmentation functions to multi-hadron fragmentation functions.  The joint track functions provide the general class of non-perturbative matrix elements necessary to compute general correlation functions $\langle \cE_{R_1}(n_1) \cdots \cE_{R_k}(n_k) \rangle$ between energy flux on particles of different quantum numbers. As such, they considerably extend the class of jet substructure measurements that are possible, and extend the application of perturbation theory to more general jet substructure observables.  We can view this space of observables schematically as in \Fig{fig:detectorspace}. Within this general space are the observables expressed in terms of well defined detectors, which can be understood using the joint track function approach, and which significantly extends the space of infrared and collinear safe observables.

An outline of this paper is as follows. In \Sec{sec:joint_tracks} we review the standard track function formalism, and extend it to define joint track functions. In \Sec{sec:joint_renorm} we study the renormalization group evolution of joint track functions. In \Sec{sec:fact_theorem}, to illustrate the application of the joint track functions, we derive a number of factorization theorems for energy correlators that involve different moments of the joint track functions. In \Sec{sec:numeric} we present results for the joint track functions for $\pm$ charges,  extracted from the Pythia parton shower \cite{Sjostrand:2014zea,Sjostrand:2007gs}. In \Sec{sec:numeric_correlators} we use these non-perturbative inputs to provide numerical results for a variety of energy correlators incorporating charges.  We conclude in \Sec{sec:conc}.

%%%%%%%%%%%%%%%%%%%%%%
\section{Joint Track Functions: General Formalism}\label{sec:joint_tracks}
%%%%%%%%%%%%%%%%%%%%%%  

In this section, we review the field theoretic definitions of the standard track functions, and extend them to joint track functions.

%%%%%%%%%%%%%%%%%%%%%%
\subsection{Review of Track Functions}\label{sec:review_track}
%%%%%%%%%%%%%%%%%%%%%%  

Track functions were originally introduced \cite{Chang:2013rca,Chang:2013iba} to enable the calculation of jet substructure observables on tracks, i.e. charged particles. In this case they are non-perturbative matrix elements that measure the fraction of energy carried by charged particles arising from the fragmentation of a quark or gluon. However, the use of charged particles is merely a specific example, and the formalism and renormalization group evolution of the matrix elements applies when the restriction to charged particles is replaced by any subset of hadrons specified by some quantum number, $R$. In this case, the track functions for quarks and gluons are defined as \cite{Chang:2013rca,Chang:2013iba} 
%%%
\begin{align} \label{T_def}
T_q(x)&=\!\int\! \df y^+ \df ^{d-2} y_\perp e^{ik^- y^+/2} \sum_X \delta \biggl( x\!-\!\frac{P_R^-}{k^-}\biggr)  \frac{1}{2N_c}
\text{tr} \biggl[  \frac{\gamma^-}{2} \langle 0| \psi(y^+,0, y_\perp)|X \rangle \langle X|\bar \psi(0) | 0 \rangle \biggr]\,,
 \\
T_g(x)&=\!\int\! \df y^+ \df^{d-2} y_\perp e^{ik^- y^+/2} \sum_X \delta \biggl( x\!-\!\frac{P_R^-}{k^-}\biggr) \frac{-1}{(d\!-\!2)(N_c^2\!-\!1)k^-}
 \langle 0|G^a_{- \lambda}(y^+,0,y_\perp)|X\rangle \langle X|G^{\lambda,a}_- (0)|0\rangle. 
\nn \end{align} 
%%%
Here $P_R^-$ denotes the longitudinal momentum fraction carried by hadrons with property $R$. We have suppressed gauge links in writing the definitions of the track functions, for simplicity. Note that the track functions measure the \emph{total} energy fraction of all hadrons with property $R$, as opposed to fragmentation functions, which measure the energy fraction of individual hadrons. This enables the track functions to describe multi-point correlations, $\langle \cE_R(n_1) \cdots \cE_R(n_k) \rangle$, as relevant for jet substructure.

Detailed discussions of the properties of track functions can be found in \cite{Chang:2013rca,Chang:2013iba,Chen:2022pdu,Chen:2022muj,Jaarsma:2022kdd,Li:2021zcf}, and we will review them only as they become necessary for our discussion. For our purposes, it will be important that the track functions satisfy a normalization condition
%%%
\begin{align}\label{T0}
T_a(0,\mu)=1\,.
\end{align}
%%%
We will often work in terms of the moments of the track functions, which we will denote
%%%
\begin{align} \label{eq:T_mom}
T_a(n,\mu)=\int \limits_0^1 \df x~ x^n~ T_a (x,\mu)\,.
\end{align}
%%%
These moments are of particular interest, since they appear in factorization theorems for the energy correlators \cite{Chen:2020vvp,Li:2021zcf,Jaarsma:2022kdd,Chen:2022pdu,Chen:2022muj}.

\begin{figure}
\begin{center}
      \subfloat[]{
\includegraphics[scale=0.2]{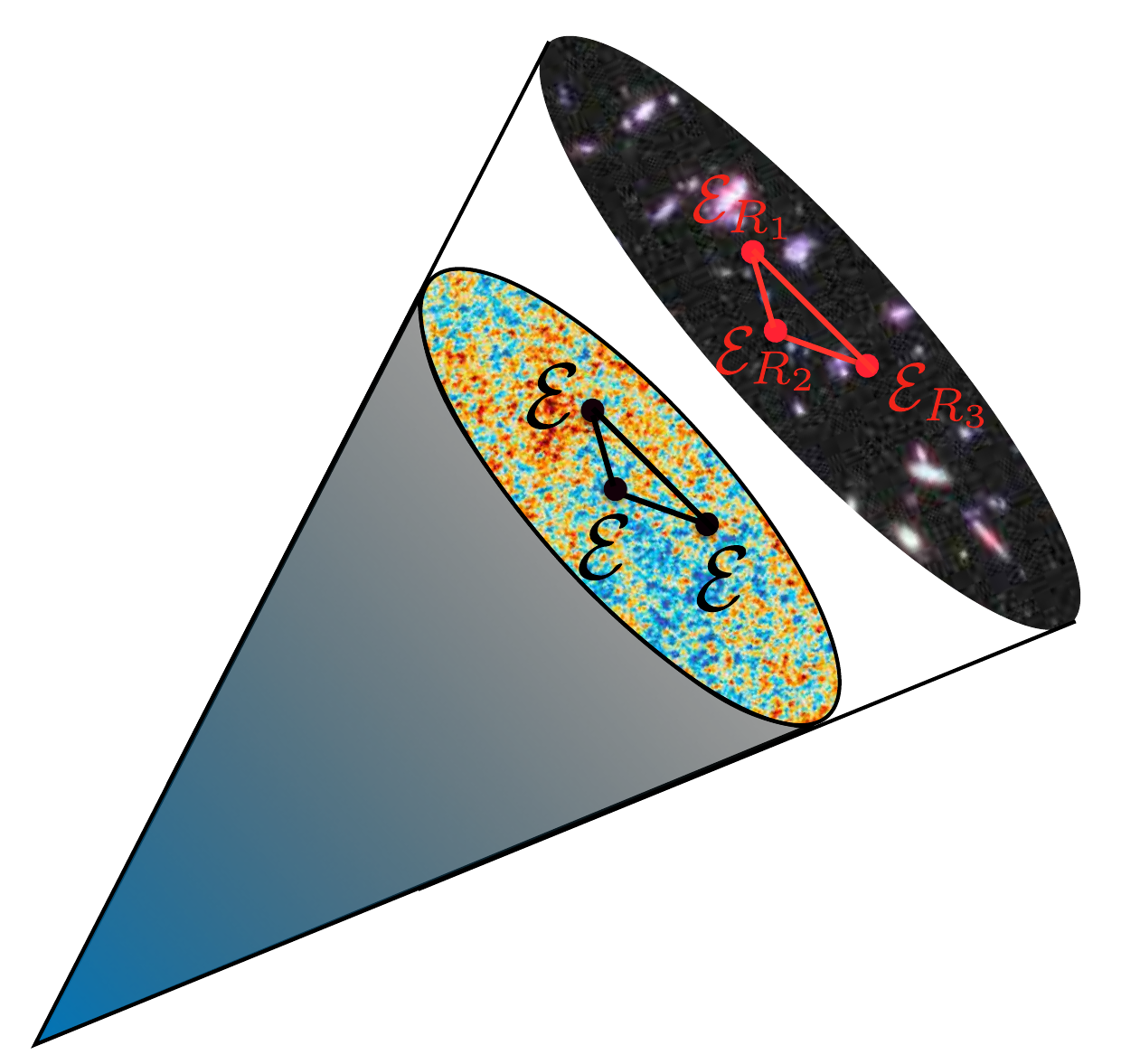} 
\label{fig:2stage}
}
      \subfloat[]{
\includegraphics[scale=0.2]{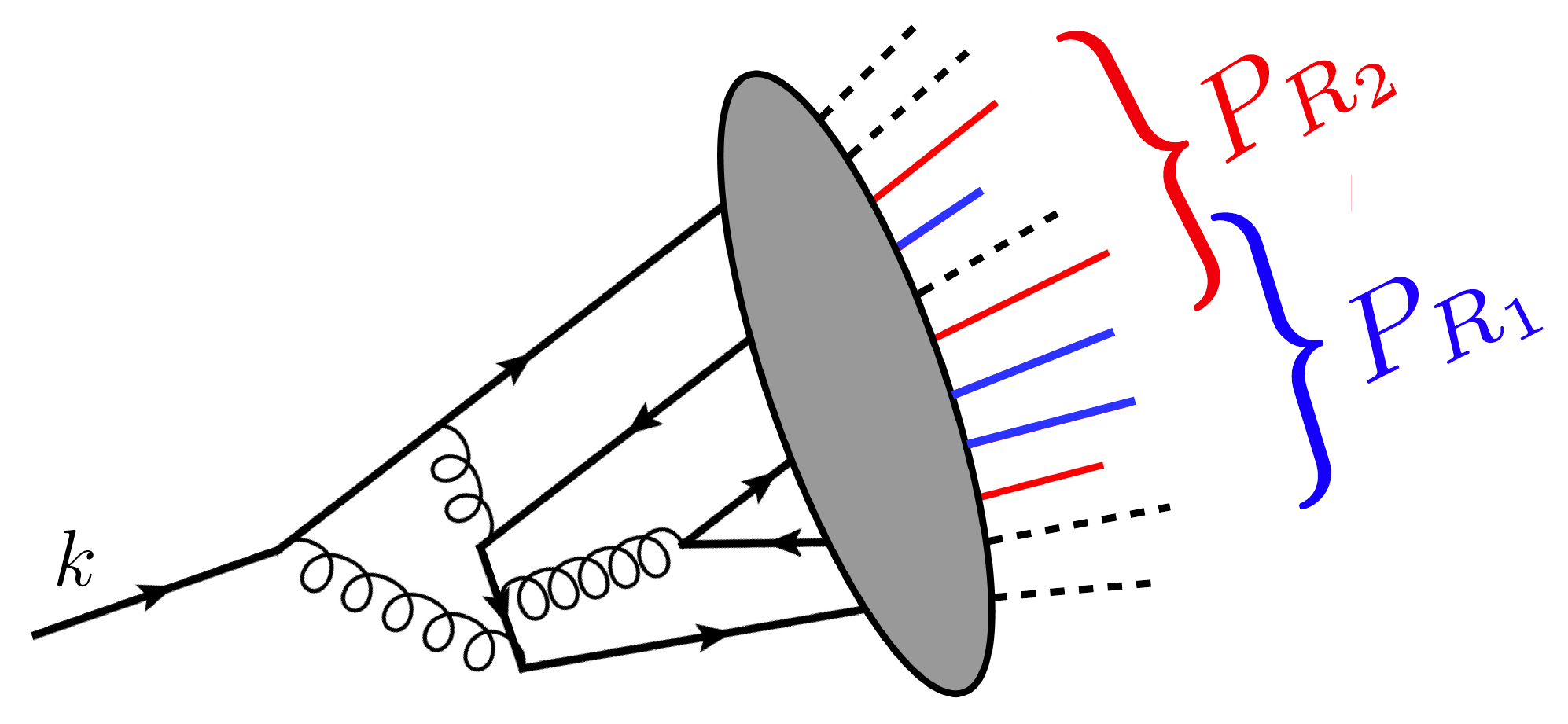} 
\label{fig:jt}
}
\end{center}
\caption{ (a)
For energy correlator observables, joint track functions provide the non-perturbative matching coefficients between parton level detectors, illustrated here on the CMB, and hadron level detectors, illustrated here on large scale structure. (b) An illustration of the quark joint track function, measuring the energy fraction carried by two distinct subsets of hadrons, $R_1$ and $R_2$, on a fragmenting quark.}
\label{fig:jt_total}
\end{figure}

%%%%%%%%%%%%%%%%%%%%
\subsection{The Joint Track Function}
%%%%%%%%%%%%%%%%%%%%

As emphasized above, track functions provide the non-perturbative matrix elements necessary to describe energy correlations on energy flux specified by a single quantum number. For example, $\langle \cE_+(n_1) \cE_+(n_2) \rangle$. While this has been extremely successful, they are inadequate to measure correlation functions on energy flux where different detectors measure the energy flux on different quantum numbers. A prime example being the physically motivated case of $\langle \cE_+(n_1) \cE_-(n_2) \rangle$.  To be able to compute such correlation functions, we must extend the track function formalism to ``joint track functions", which are a class of matrix elements necessary to compute generic energy correlation functions $\langle \cE_{R_1}(n_1) \cdots \cE_{R_k}(n_k) \rangle$, where each detector measures the energy flux on states specified by the quantum number $R_k$. As compared to the case of track functions, the $R_k$ can now generally be distinct. 

We begin with the most generic definition before restricting to specific cases of interest.  Let $R_1, \cdots, R_k$ denote a set of quantum numbers on which we would like to study the energy flux. We can define a joint track function that simultaneously measures the energy fraction from a fragmenting quark or gluon into each of these quantum numbers. We term this a ``joint track function".  The operator definition for the quark and gluon joint track functions are
%%%
{
\begin{align}
T^{R_1 \cdots R_k}_q(x_{R_1}, \cdots, x_{R_n} )&=\!\int\! \df y^+ \df ^{d-2} y_\perp e^{ik^- y^+/2} \sum_X  \delta \biggl( x_{R_1}\!-\!\frac{P_{R_1}^-}{k^-}\biggr) \cdots \delta \biggl( x_{R_n}\!-\!\frac{P_{R_n}^-}{k^-}\biggr)\nn \\
&\cdot  \frac{1}{2N_c}
\text{tr} \biggl[  \frac{\gamma^-}{2} \langle 0| \psi(y^+,0, y_\perp)|X \rangle \langle X|\bar \psi(0) | 0 \rangle \biggr]\,, \nn
 \\
T^{R_1 \cdots R_k}_g(x_{R_1}, \cdots, x_{R_k})&=\!\int\! \df y^+ \df^{d-2} y_\perp e^{ik^- y^+/2} \sum_X \delta \biggl( x_{R_1}\!-\!\frac{P_{R_1}^-}{k^-}\biggr) \cdots \delta \biggl( x_{R_n}\!-\!\frac{P_{R_n}^-}{k^-}\biggr)\nn \\
&\cdot \frac{-1}{(d\!-\!2)(N_c^2\!-\!1)k^-}
 \langle 0|G^a_{- \lambda}(y^+,0,y_\perp)|X\rangle \langle X|G^{\lambda,a}_- (0)|0\rangle. 
\nn \end{align} 
}
As with the definition of the track function given above, we have suppressed the gauge links for simplicity.
We see that these joint track functions are nearly identical to the standard track functions, except with additional measurement functions which measure multiple $P_{R_i}^-$, namely the longitudinal momentum fraction carried by hadrons with property $R_i$. An illustration of the joint track function is shown in \Fig{fig:jt} for the case of a joint quark track function with two subsets of hadrons, $R_1$ and $R_2$. The extension from track functions to joint track functions can be viewed much like the extension from single hadron fragmentation functions to multi-hadron fragmentation functions.

By itself, this definition is not particularly useful. Indeed, it defines a complicated multi-variable non-perturbative function. However, the key recent advance in jet substructure has been the understanding of how to ask questions about jets so as to be sensitive only to specific moments of track functions \cite{Chen:2020vvp}, or in this case joint track functions. Therefore while standard jet shape observables require complete non-perturbative track \emph{functions}, we will see that the correlation functions $\langle \cE_{R_1}(n_1) \cdots \cE_{R_k}(n_k) \rangle$ require only moments, i.e. \emph{numbers}. This makes the incorporation of non-perturbative information practical, and as we will see, will enable us to draw interesting conclusions about more general correlation functions.

Much as for the standard track functions, it will be useful to work with moments of the joint track functions, which will appear in the calculation of the charged energy correlators, also simply referred to as charged correlators. We will denote the multiple moments of the joint track function as
\begin{align}
T^{R_1 \cdots R_k}(n_1, \cdots,n_k;\mu) =& \int dx_1\cdots dx_k \, x_1^{n_1}\, x_k^{n_k}\,T_{R_1 \cdots R_k}(x_1,\cdots,x_k;\mu)\,.
\label{eq:jtmom}
\end{align}
These obey a variety of sum rules. First, they obey an overall normalization
\begin{align}
T^{R_1 \cdots R_k}(0, \cdots, 0;\mu) =& \int dx_1\cdots dx_k \, T_{R_1 \cdots R_k}(x_1,\cdots,x_k;\mu)=1\,.
\end{align}
Second, integrating over any subset of variables, reduces the joint track function to a joint track function over the remaining variables. Let $S$ denote the set of variables, and $S_1 \subset S$ be a subset of variables. Then, we have the sum rule
\begin{align}
\int \prod\limits_{i\in S_1} dx_i  \, T_{R_1 \cdots R_k}(x_1,\cdots,x_k;\mu)=T_{\prod\limits_{i\in S\backslash S_1} R_i }(x_i|i\in S\backslash S_1;\mu)\,.
\end{align}
Note that the quantum numbers do not have to be mutually exclusive for this to apply. For example, the quantum numbers could be strangeness and charge.

While we have so far kept our discussion fully general, a particular example of the joint track function of experimental relevance, is the joint track function that measures both positive and negative charge, $T_i^{+-}(x,y)$. This is directly applicable to jet substructure measurements at collider experiments, which can measure electromagnetic charge, and it extends the study of energy correlation functions to correlators involving both $\cE_+(n_1)$ and  $\cE_-(n_1)$. Since we expect this matrix element to be particularly relevant for future studies, we give it the shorthand name $T_i^{+-}(x,y)\,,$ where $x$ and $y$ represent the momentum fraction of $+$ and $-$ tracks, respectively. This function can be projected onto the usual track functions using the following sum rules
\begin{align}
\label{eq:1dfulltrack}
T_i(z) &= \int dx\,dy\, T_i^{+-}(x,y)\, \delta(z-x-y)\,,\\
T_i^+(x) &=  \int dy\, T_i^{+-}(x,y)\,,\\
T_i^-(y) &=  \int dx\, T_i^{+-}(x,y)\,.
\label{eq:1d-track}
\end{align}
Here, $T_i(z)$ is the standard track function that measures the energy flux on all charged hadrons, while $T_i^+$ and $T_i^-$ measure the energy flux on positve and negative hadrons, respectively.

We believe that joint track functions provide the general framework for studying the factorization of energy correlators on subsets of hadrons, $\langle \cE_{R_1}(n_1) \cdots \cE_{R_k}(n_k) \rangle$, and we will show that they enable qualitatively new calculations extending the boundaries of perturbation theory. We therefore believe that they will be a powerful tool for better understanding the parton to hadron transition.

%%%%%%%%%%%%%%%%%%%%%%
\section{Renormalization Group Analysis}\label{sec:joint_renorm}
%%%%%%%%%%%%%%%%%%%%%%  

Much like the standard track functions, or fragmentation functions, the joint track functions are non-perturbative, but their renormalization group evolution can be computed in perturbation theory. While the timelike DGLAP evolution \cite{Dokshitzer:1977sg,Gribov:1972ri,Altarelli:1977zs} of the fragmentation functions is well known, the RG equations describing the evolution of the track functions were only recently computed beyond the leading order  \cite{Chen:2022muj,Chen:2022pdu,Jaarsma:2022kdd,Li:2021zcf}. A key aspect of this work was the derivation of a general set of collinear evolution equations \cite{Chen:2022muj,Chen:2022pdu}. Since these evolution equations describe the full momentum dependence of collinear evolution, they also describe the renormalization group evolution of the joint track functions. In this section we begin by reviewing the renormalization group evolution of the track functions, both in momentum fraction space, following \cite{Chen:2022muj,Chen:2022pdu}, and in moment space, following \cite{Jaarsma:2022kdd,Li:2021zcf}, and then generalize it to the joint track functions.

%%%%%%%%%%%%%%%%%%%%%%
\subsection{$x$-Space}\label{sec:pos_space}
%%%%%%%%%%%%%%%%%%%%%%  

The renormalization group evolution of the track functions was systematically understood in \cite{Jaarsma:2022kdd,Li:2021zcf,Chen:2022muj,Chen:2022pdu}. It is expressed in terms of a set of perturbatively calculable kernels, $K_{i\to i_1i_2, \cdots, i_n}$, which describe the evolution of a track function $T_i$ to a product of track functions $T_{i_1} \cdots T_{i_n}$. Up to NLO, one can have a transition into at most three track functions, and the evolution takes the form
%%%
\begin{align}
 \!\!\!\frac{\df }{\df \ln\mu^2}T_i(x) &= a_s \Bigl(
 K^{(0)}_{i\to i} T_i(x)  +
[K^{(0)}_{i\to i_1i_2}\otimes T_{i_1}  T_{i_2}](x) \Bigr)
 \nn \\ & \quad
 +
a_s^2\Bigl(
K^{(1)}_{i\to i} T_i(x) 
+
[K^{(1)}_{i\to i_1i_2}\otimes T_{i_1}  T_{i_2}](x)
 \nn \\
  & \quad  \!+\! [K^{(1)}_{i\to i_1i_2i_3}\otimes T_{i_1} T_{i_2} T_{i_3}](x)
  \Bigr) \!+\! \mathcal{O}(a_s^3)\,,
\end{align}
%%%
where we have used the reduced coupling $a_s=\alpha_s/(4\pi)$.
The explicit kernels can be found in \cite{Chen:2022pdu,Chen:2022muj}. In terms of the moments of the track functions, this gives the evolution equation
%%%
\begin{align} \label{T_evo_n}
  \frac{\df}{\df \ln \mu^2} T_a(n) &= \sum_N \sum_{\{a_f\}} \sum_{\{m_f\}} \ga_{a \to \{a_f\}}(\{m_f\})\,  \prod_{i=1}^N T_{a_i}(m_i,\mu) \,,
\end{align}
%%%
where results for the anomalous dimensions can be found in \cite{Jaarsma:2022kdd,Li:2021zcf}.

The renormalization group evolution of the joint track function is a simple extension of that for the track function, and indeed, can be computed from the same kernels. To keep the notation compact, we first present the result for the joint track function with two arguments. The extension to more general joint track functions then follows straightforwardly. The general evolution equation for the joint track functions can be written in terms of the kernels $K_{i\to i_1i_2, \cdots, i_n}$, and differs only in the convolution structure. Up to NLO, we have
\begin{align}
\label{eq:jointevol}
\frac{\mathrm{d}}{\mathrm{d} \ln \mu^2} T_i^{+-}(x,y)&=a_s  {\left[K_{i \rightarrow i}^{(0)} T_i^{+-}(x,y)+K_{i \rightarrow i_1 i_2}^{(0)} \otimes T_{i_1}^{+-} T_{i_2}^{+-}(x,y)\right] } \\
& \hspace{-1.5cm}+a_s^2\left[K_{i \rightarrow i}^{(1)} T_i^{+-}(x,y)+K_{i \rightarrow i_1 i_2}^{(1)} \otimes T_{i_1}^{+-} T_{i_2}^{+-}(x,y)+K_{i \rightarrow i_1 i_2 i_3}^{(1)} \otimes T_{i_1}^{+-} T_{i_2}^{+-} T_{i_3}^{+-}(x,y)\right]\,. \nn
\end{align}
The convolutions are now defined as
\begin{align}
&K_{i \rightarrow i_1 i_2} \otimes T_{i_1}^{+-} T_{i_2}^{+-}(x,y)=\int_0^1 dx_1\, dx_2 \,dy_1 \,dy_2\,T_{i_1}^{+-}\left(x_1,y_1\right) T_{i_2}^{+-}\left(x_2,y_2\right)\\
&\times \int_0^1 dz_1\, dz_2 \delta\left(1-z_1-z_2\right) \delta\left(x-z_1 x_1-z_2 x_2\right)\delta\left(y-z_1 y_1-z_2 y_2\right) K_{i \rightarrow i_1 i_2}\left(z_1, z_2\right)\,,\nn
\end{align}
for the $1\to 2$ kernel, and 
\begin{align}
&K_{i \rightarrow i_1 i_2 i_3} \otimes T_{i_1}^{+-} T_{i_2}^{+-}T_{i_3}^{+-}(x,y)= \hspace{-0.15cm}\int_0^1\hspace{-0.15cm} dx_1\, dx_2 \,dy_1 \,dy_2\,T_{i_1}^{+-}\left(x_1,y_1\right) T_{i_2}^{+-}\left(x_2,y_2\right) T_{i_3}^{+-}\left(x_3,y_3\right)\\
&\cdot \int_0^1 dz_1\, dz_2 \delta\left(1-z_1-z_2-z_3\right) \delta\left(x-z_1 x_1-z_2 x_2-z_3 x_3\right)\delta\left(y-z_1 y_1-z_2 y_2-z_3 y_3\right) K_{i \rightarrow i_1 i_2}\left(z_1, z_2,z_3\right)\,,\nn
\end{align}
for the $1\to 3$ kernel. The extension to even higher dimensional joint track functions can be straightforwardly achieved by incorporating additional variables into the convolutions, similar to the transition from a 1-dimensional track function to a 2-dimensional joint track function.

%%%%%%%%%%%%%%%%%%%%%%
\subsection{Moment Space and Asymptotics of RG Flows}\label{sec:moments_asym}
%%%%%%%%%%%%%%%%%%%%%%  

For many purposes it is more convenient to work in terms of the moments of the joint track functions. Their organization also follows closely that of the moments of the standard track functions. We again illustrate with the case of $T_i^{+-}$. We can define
\begin{align}
T_i^{+-}(N,M;\mu) =& \int dx\,dy\, x^N\, y^M\,T_i^{+-}(x,y;\mu)\,,\\
T_i^{\pm}(N;\mu) =&  \int dx\, x^N\,T_i^{\pm}(x;\mu)\,.
\end{align}
Note that $T_i^{+-}(N,0;\mu) = T_i^+(N;\mu)$ and $T_i^{+-}(0,N;\mu) = T_i^-(N;\mu)$.

In ~\cite{Jaarsma:2022kdd,Li:2021zcf}, shift symmetry was used to simplify the structure of the RG equations. It was shown that the shift symmetry uniquely fixes the RG evolution of the first three moments of the track functions in terms of the DGLAP anomalous dimensions, up to terms proportional to $\Delta_i(\mu)=T_i(1;\mu)-T_g(1;\mu)$. This is a particularly convenient organization in QCD, since $\Delta\ll 1$ \cite{Jaarsma:2022kdd}. We will express the renormalization group evolution in terms of the timelike splitting functions, $P_{ij}(z)$, and their moments.   We expand the splitting function perturbatively as
\begin{align}
P_{ij}(z)=\sum_{L=0}^\infty \Bigl(\frac{\alpha_s}{4\pi}\Bigr)^{L+1}P_{ij}^{(L)}(z)\,.
\end{align}
The timelike splitting function is known to NNLO  \cite{Stratmann:1996hn,Mitov:2006ic,Moch:2007tx,Almasy:2011eq,Chen:2020uvt}. The Mellin moments of the timelike splitting functions are given by
\begin{align}
\gamma_{ij}^{(L)}(k)=-\int_0^1\df z\ z^{k-1}P_{ij}^{(L)}(z)\ .
\end{align}
Using shift symmetry, the RG equations for the first three central moments of the standard track functions are fixed to be
\begin{align}
\label{eq:trackshift1}
\frac{\mathrm{d}}{\mathrm{d} \ln \mu^2} \Delta_i(\mu) & =-\left[\gamma_{g g}(2)+\gamma_{q q}(2)\right] \Delta_{i}(\mu), \\
\frac{\mathrm{d}}{\mathrm{d} \ln \mu^2} \vec{\sigma}(2;\mu) & =-\hat{\gamma}(3) \vec{\sigma}(2;\mu)+\vec{\gamma}_{\Delta^2_{ij}} \Delta_i(\mu)\Delta_j(\mu), \\
\frac{\mathrm{d}}{\mathrm{d} \ln \mu^2} \vec{\sigma}(3;\mu) & =-\hat{\gamma}(4) \vec{\sigma}(3;\mu)+\hat{\gamma}_{\sigma_2 \Delta_i} \vec{\sigma}(2;\mu) \Delta_i(\mu)+\vec{\gamma}_{\Delta_{ijk}^3} \Delta_{i}(\mu) \Delta_{j}(\mu) \Delta_{k}(\mu),
\label{eq:trackshift3}
\end{align}
where $\sigma_i(2;\mu)=T_i(2;\mu)-T_i(1;\mu)^2$, $\sigma_i(3;\mu)=T_i(3;\mu)-3 T_i(2;\mu) T_i(1;\mu)+2 T_i(1;\mu)^3$, and $\Delta_i(\mu)=T_i(1;\mu)-T_g(1;\mu)$. These $\sigma_i(n)$ are shift-invariant central moments and these results can be generalized to even higher degree central moments. For convenience, we define the vector $\vec{\sigma}(m;\mu)=\left(\sigma_u(m;\mu),\sigma_{\bar{u}}(m;\mu),\sigma_{d}(m;\mu), ... , \sigma_g(m;\mu)\right)$. Additionally, the matrices of timelike twist-$2$ spin-$J$ anomalous dimensions are denoted as $\hat{\gamma}(J)$, while explicit expressions for the remaining anomalous dimensions can be found in the literature up to NLO~\cite{Jaarsma:2022kdd,Li:2021zcf}. 
Beyond the third moment, genuinely new terms appear that are not related to DGLAP, and are not suppressed by $\Delta$. These correspond to mixings between $\sigma_4$ and $\sigma_2 \sigma_2$, for example.

In the case of the joint track function, $T_i^{+-}(x,y;\mu)$, we can similarly define shift invariant moments
\bea
\sigma_i^{+-}(n,m;\mu) =\int_0^1 dx dy\, (x-T^+(1;\mu))^n\,(y-T^+(1;\mu))^m\, T_i^{+-}(x,y;\mu)\,.
\eea
For low moments these are familiar objects, which will also provide us some intuition for the behavior of their evolution. For example, for $n,m=1$, we have simply the covariance
\bea
\sigma_i^{+-}(1,1;\mu)=T_i^{+-}(1,1;\mu)-T_i^{+}(1;\mu)T_i^{-}(1;\mu) = \mathrm{E}[X Y] - \mathrm{E}[X]E[Y] = \mathrm{cov}(X,Y)\,,
\eea
where we have used $X$ and $Y$ as the random variables for the $+$ and $-$ charge distributions, respectively.

For the case of the joint track functions, we are similarly able to show that the first three moments ($n+m \leq 3$) are also entirely fixed in terms of the DGLAP anomalous dimensions, up to terms proportional to $\Delta$. For $n+m=1$, the unique object is $\Delta^{\pm}$, which satisfies the RG evolution equation
\begin{align}
\frac{\mathrm{d}}{\mathrm{d} \ln \mu^2} \Delta_i^{\pm}(\mu) & =-\left[\gamma_{g g}(2)+\gamma_{q q}(2)\right] \Delta_i^{\pm}(\mu)\,.
\end{align}
At $n+m=2$, the only new object that does not reduce to standard track functions by sum rules, is $\vec{\sigma}^{+-}(1,1;\mu)$, which satisfies the RG equation
\begin{align}
\frac{\mathrm{d}}{\mathrm{d} \ln \mu^2} \vec{\sigma}^{+-}(1,1;\mu) & =-\hat{\gamma}(3) \vec{\sigma}^{+-}(1,1;\mu)+\frac{\vec{\gamma}_{\Delta^2_{ij}}}{2} \left(\Delta_i^+(\mu) \Delta_j^-(\mu)+\Delta_i^-(\mu) \Delta_j^+(\mu)\right)\,.
\end{align}
At $n+m=3$ the only new object, after taking into account symmetries and sum rules, is $\vec{\sigma}^{+-}(1,2;\mu)$, which satisfies the RG equation
\begin{align}
\frac{\mathrm{d}}{\mathrm{d} \ln \mu^2} \left(\vec{\sigma}^{+-}(1,2;\mu)+\vec{\sigma}^{+-}(2,1;\mu)\right) & =-\hat{\gamma}(4) \left(\vec{\sigma}^{+-}(1,2;\mu)+\vec{\sigma}^{+-}(2,1;\mu)\right) \\
&\hspace{-4cm}+\frac{\hat{\gamma}_{\sigma_2 \Delta_i}}{3} \left[\vec{\sigma}^+(2;\mu) \Delta_i^-(\mu)+\vec{\sigma}^-(2;\mu) \Delta_i^+(\mu)+ 2\vec{\sigma}^{+-}(1,1;\mu) \left(\Delta_i^+(\mu)+\Delta_i^-(\mu)\right)\right]\nonumber\\
&\hspace{-4cm}+\frac{\vec{\gamma}_{\Delta_{ijk}^3}}{3} \bigg[\Delta_i^-(\mu)\Delta_j^+(\mu)\Delta_k^+(\mu)+\Delta_i^+(\mu)\Delta_j^-(\mu)\Delta_k^+(\mu)+\Delta_i^+(\mu)\Delta_j^+(\mu)\Delta_k^-(\mu)\nonumber\\
&\hspace{-4cm}+\Delta_i^-(\mu)\Delta_j^-(\mu)\Delta_k^+(\mu)+\Delta_i^-(\mu)\Delta_j^+(\mu)\Delta_k^-(\mu)+\Delta_i^+(\mu)\Delta_j^-(\mu)\Delta_k^-(\mu)\bigg]\,. \nn
\end{align}
These equations are particularly interesting due to the fact that $\Delta \ll 1$ in QCD, as highlighted in \cite{Jaarsma:2022kdd}. Therefore, they tell us that to a very good approximation, $\vec{\sigma}^{+-}(1,1;\mu)$ evolves with $\hat \gamma(3)$ and $\vec{\sigma}^{+-}(2,1;\mu)$ evolves with $\hat \gamma(4)$.  This is quite remarkable, since it allows us to understand certain universal properties of the renormalization group flow in the space of moments of the joint track functions. Although the detailed structure of the track functions require non-perturbative inputs, the renormalization group allows one to make certain qualitative statements about the structure of the hadronization transition by studying fixed points in the space of non-perturbative matrix elements characterizing the transition. For example in the case of standard track functions, the structure of anomalous dimensions allows one to show that higher central moments are rapidly damped as one evolves into the UV \cite{Jaarsma:2022kdd}. It is therefore interesting to understand what general statements can be made about the joint track functions, which provides information about how correlations evolve as a function of scale.

It is known that in unitary theories, the eigenvalues of $\hat \gamma(3)$ are positive \cite{Nachtmann:1973mr,Komargodski:2016gci}. Therefore, from the RG evolution, we can conclude that independent of the non-perturbative input, the only fixed point for $\vec{\sigma}^{+-}(1,1;\mu)$ is at $\vec{\sigma}^{+-}(1,1;\mu)\to 0$, i.e. of vanishing covariance. We will see from parton shower Monte Carlo simulations that this begins to happen at relatively low energies. We find this to be an interesting example of the form of statement that can be made using renormalization group arguments. Note that even if $\Delta$ were sizable in QCD, its RG equation ensures that it would first be driven to zero, and then the same conclusion holds. More generally, much like for the standard track functions, all higher central moments will also be driven to zero. This can be seen from the fact that all lower moment terms appearing in the non-linear mixing term will be driven to zero, starting from the positivity of $\left[\gamma_{g g}(2)+\gamma_{q q}(2)\right]$ quickly decays $\Delta_i^\pm(\mu)\to 0$. In particular, we observe that the covariance of joint track functions are driven to zero as they evolve to higher scale and also eventually becomes a delta function in the two-dimensional space.

An interesting feature of the multivariable nature of the joint track functions is that there are a variety of different moments that one can consider. In addition to the multivariable central moments, we will also find that ``asymmetries" of the joint track function appear in factorization theorems for energy correlators. We define the following moments of the joint track function
\begin{align}
T_i^{\mathrm{asy}}(N;\mu)=&\int_0^1 \,dx\, dy\, (x-y)^N\, T_i^{+-}(x,y;\mu)\,.
\end{align}
The asymmetry track function obeys the usual renormalization group equations of the 1-dimensional track functions
\begin{align}
 \!\!\!\frac{\df }{\df \ln\mu^2}T^{\mathrm{asy}}_i(x) &= a_s \Bigl(
 K^{(0)}_{i\to i} T^{\mathrm{asy}}_i(x)  +
[K^{(0)}_{i\to i_1i_2}\otimes T^{\mathrm{asy}}_{i_1}  T^{\mathrm{asy}}_{i_2}](x) \Bigr)
 \nn \\ & \quad
 +
a_s^2\Bigl(
K^{(1)}_{i\to i} T^{\mathrm{asy}}_i(x) 
+
[K^{(1)}_{i\to i_1i_2}\otimes T^{\mathrm{asy}}_{i_1}  T^{\mathrm{asy}}_{i_2}](x)
 \nn \\
  & \quad  \!+\! [K^{(1)}_{i\to i_1i_2i_3}\otimes T^{\mathrm{asy}}_{i_1} T^{\mathrm{asy}}_{i_2} T^{\mathrm{asy}}_{i_3}](x)
  \Bigr) \!+\! \mathcal{O}(a_s^3)\,,
\end{align}
and has a domain between $-1<z<1$. It also satisfies the usual shift symmetry and the RG equations for the first three central moments can be written as
\begin{align}
\label{eq:asytrackshift1}
\frac{\mathrm{d}}{\mathrm{d} \ln \mu^2} \Delta^{\mathrm{asy}}_i(\mu) & =-\left[\gamma_{g g}(2)+\gamma_{q q}(2)\right] \Delta^{\mathrm{asy}}_{i}(\mu), \\
\frac{\mathrm{d}}{\mathrm{d} \ln \mu^2} \vec{\sigma}^{\mathrm{asy}}(2;\mu) & =-\hat{\gamma}(3) \vec{\sigma}^{\mathrm{asy}}(2;\mu)+\vec{\gamma}_{\Delta^2_{ij}} \Delta^{\mathrm{asy}}_i(\mu)\Delta^{\mathrm{asy}}_j(\mu), \\
\frac{\mathrm{d}}{\mathrm{d} \ln \mu^2} \vec{\sigma}^{\mathrm{asy}}(3;\mu) & =-\hat{\gamma}(4) \vec{\sigma}^{\mathrm{asy}}(3;\mu)+\hat{\gamma}_{\sigma_2 \Delta_i} \vec{\sigma}^{\mathrm{asy}}(2;\mu) \Delta^{\mathrm{asy}}_i(\mu)+\vec{\gamma}_{\Delta_{ijk}^3} \Delta^{\mathrm{asy}}_{i}(\mu) \Delta^{\mathrm{asy}}_{j}(\mu) \Delta^{\mathrm{asy}}_{k}(\mu),
\label{eq:asytrackshift3}
\end{align}
where $\sigma^{\mathrm{asy}}_i(2;\mu)=T^{\mathrm{asy}}_i(2;\mu)-T^{\mathrm{asy}}_i(1;\mu)^2$, $\sigma^{\mathrm{asy}}_i(3;\mu)=T^{\mathrm{asy}}_i(3;\mu)-3 T^{\mathrm{asy}}_i(2;\mu) T^{\mathrm{asy}}_i(1;\mu)+2 T^{\mathrm{asy}}_i(1;\mu)^3$, and $\Delta^{\mathrm{asy}}_i(\mu)=T^{\mathrm{asy}}_i(1;\mu)-T^{\mathrm{asy}}_g(1;\mu)$. At some general scale, $T_g^{\mathrm{asy}}\approx 0, T_u^{\mathrm{asy}}>0,$ and $T_d^{\mathrm{asy}}<0$. Therefore, the first moment of the quark asymmetry track functions will approach zero from different sides for up and down-types. Then higher moments will also be driven to zero in the UV as usual, giving a delta function at zero charge asymmetry in the UV.

%%%%%%%%%%%%%%%%%%%%%%
\section{Factorization Theorems Involving Joint Track Functions}\label{sec:fact_theorem}
%%%%%%%%%%%%%%%%%%%%%%  

Having introduced the joint track functions, we now show how different combinations of their moments  enter into factorization theorems for generalized energy correlators involving electromagnetic charge. A particularly interesting feature arising from the fact that due to the fact that the joint track functions are multi-variable functions, a wider variety of different moments exist, and these appear in the description of different energy correlator observables. Higher moments of the joint track functions only appear in contact terms of energy correlators, and hence in the resummation in the small angle limit. For widely separated detectors, only single moments of the joint track functions appear, which always reduce to first moments of standard track functions. To focus on the contributions from the joint track functions, we will study generalizations of projected energy correlators \cite{Chen:2020vvp}, where different higher moments of the joint track functions appear.

%%%%%%%%%%%%%%%%%%%%%%%%%%%%%%%%%%%
\subsection{$\mathcal{E}_{\pm}$ Correlators}
%%%%%%%%%%%%%%%%%%%%%%%%%%%%%%%%%%%

First, we define $(N,M)$-point charge correlators with $N$ and $M$ detectors on positive and negative tracks, respectively. They are explicitly denoted as
\bea
\langle \underbrace{\mathcal{E}_+ \cdots \mathcal{E}_+}_{N \text{ times}} \underbrace{\mathcal{E}_- \cdots \mathcal{E}_-}_{M \text{ times}} \rangle\,.
\eea
We focus on the projected multi-point charge correlators by integrating out all the information about the shape while keeping the length of the longest side, $R_L$, measured. This is a natural generalization of the standard projected energy correlators \cite{Chen:2020vvp,Lee:2022ige}. In the small-angle limit, the cumulant of the projected $(N,M)$-point charge correlators $\Sigma_{+-}^{[N,M]}\left(R_L, p_T^2, \mu\right)$ then factorizes into a hard function $\vec{H}\left(x, p_T^2, \mu\right)$, which describes the production of the collinear source, and the $(N,M)$-point charge energy correlator jet function, $\vec{J}_{+-}^{[N,M]}\left(R_L, x, \mu \right)$, which describes the $R_L$ dependence of the observable. This is a natural extension of the factorization theorems of \cite{Dixon:2019uzg,Chen:2020vvp}
\begin{align}
& \Sigma^{[N,M]}_{+-}\left(R_L, p_T^2, \mu\right) =\int_0^1 d x~ x^{N+M} \vec{J}_{+-}^{[N,M]}\left(R_L, x,  \mu\right) \cdot \vec{H}\left(x, p_T^2, \mu\right)\,.
\end{align}
By renormalization group consistency, the general charge energy correlator jet function obeys the renormalization group evolution equation
\bea
\frac{d \vec{J}_{+-}^{[N,M]}\left(\ln \frac{p_T^2 R_L^2}{\mu^2}\right)}{d \ln \mu^2}=\int_0^1 dy \, y^{N+M} \vec{J}_{+-}^{[N,M]}\left(\ln \frac{y^2 p_T^2 R_L^2}{\mu^2}\right) \cdot \widehat{P}(y)\,,
\eea
where $\widehat{P}(y)$ is the singlet timelike DGLAP splitting kernel matrix. 

We can now apply the joint track function formalism to factorize the jet function into a perturbatively calculable coefficients and the moments of the joint track function, which encodes the universal non-perturbative QCD dynamics of the observable. Most interestingly, this observable does not depend on the full functional form of the joint track function, but only the $(N,M)$ moment, as well as the lower moments that this can mix with under renormalization group evolution. We can thus write this matching relation in the abstract notation
\begin{align}
J^{+-}_k= {\bf{T^{+-}_k(N,M)}} \cdot {\bf{J^{+-}_k}}\,.
\end{align}
Here we have generalized the notation in \cite{Jaarsma:2022kdd}, and ${\bf{T^{+-}_k(N,M)}}$ denotes a vector in the space of joint track functions, consisting of all moments of joint track functions of weight $N+M$, which are precisely those that appear in the RG mixing. For example, ${\bf{T^{+-}_k(1,1)}}=(  \sigma^{+-}(1,1),\Delta^+(1) \Delta^-(1))$\,. The corresponding vector ${\bf{J^{+-}_k}}$ denotes the perturbatively calculable coefficients. This is a straightforward extension for the case of the energy correlators factorized onto standard track function.

At leading logarithmic order, this is particularly simple, since one only needs the tree level boundary condition for the jet function, which is given by
\begin{align}
\label{eq:Jbdry}
\hspace{-0.5cm}\vec{J}^{[N,M]}_{+-}\left(\mu=p_T R_L\right)&=\left(T_u^{+-}(N,M;p_T R_L),T_{\bar{u}}^{+-}(N,M;p_T R_L),T_d^{+-}(N,M;p_T R_L),... , T_g^{+-}(N,M;p_T R_L)\right)\nonumber\\
& \equiv\vec{T}^{+-}(N,M;p_T R_L) \,.
\end{align}
This allows us to straightforwardly derive the leading logarithmic scaling behavior for the projected $(N,M)$-point charge correlators, from the renormalization group evolution of the jet function
\bea
\label{eq:chargeLL}
\vec{J}^{[N,M]}_{+-,{\rm LL}}\left(\ln \frac{p_T^2 R_L^2}{\mu^2}, \mu\right)=\vec{T}^{+-}(N,M;p_T R_L)\cdot \left(\frac{\alpha_s(p_T R_L)}{\alpha_s(\mu)}\right)^{-\frac{\hat{\gamma}^{(0)}(N+M+1)}{\beta_0}}\,.
\eea
We see that this result  reduces to the standard track energy correlators for $N$ or $M=0$, as expected. More generally, it probes the  $+$ and $-$ charge distribution at the physical scale given by the EEC angle $\mu = p_T R_L$. This allows us to describe a much broader class of jet substructure observables than possible previously. Note that the overall scaling behavior is still governed by $\gamma(N+M+1)$ \cite{Konishi:1979cb,Belitsky:2013ofa,Dixon:2019uzg,Kologlu:2019mfz,Korchemsky:2019nzm,Chen:2020vvp}, but crucially, the renormalization group evolution of the non-perturbative matrix elements can lead to perturbatively calculable modifications to this scaling.

From a physics perspective, this factorization theorem allows us to isolate in a physical observable the $(N,M)$ moments of the joint track function. The scaling behavior of these correlators therefore only requires non-perturbative \emph{numbers}, as opposed to non-perturbative functions. This is the particularly interesting feature of the energy correlator based approach to jet substructure.

We are also able to describe the leading non-perturbative power corrections to our factorization formula. These are crucial for obtaining good numerical agreement with parton shower Monte Carlo programs. In addition to the simple manner in which moments of the joint track functions enter the factorization theorems for the energy correlators, another key advantage is that the functional form of the leading power correction can be predicted.  Indeed, it can be shown that the leading non-perturbative power correction is additive, and its functional form is given by \cite{Belitsky:2001ij,Korchemsky:1999kt,Korchemsky:1997sy,Korchemsky:1995zm,Korchemsky:1994is,Schindler:2023cww}  
%%%
\begin{align}
\text{EEC}(\theta)=\text{EEC}_\text{pert}(\theta)+\frac{\Lambda_1}{p_T}\frac{1}{\theta^{3}}\,,
\end{align}
%%%
which is fixed by boost invariance. Here, the classical scaling of the perturbative part is $\sim \frac{1}{\theta^2}$. For the standard energy-energy correlators case involving all hadrons, the $\Lambda_1 \sim 1$ GeV was shown to be universal nonperturbative parameter. This continues to hold for the more general correlators here, with the only difference being that the leading non-perturbative parameter $\Lambda_1$ is rescaled by appropriate moments of the joint track functions. This was first shown in \cite{yibei:2023}.

%%%%%%%%%%%%%%%%%%%%%%%%%%%%%%%%%%%
\subsection{$\mathcal{E}_Q$ Correlators}
%%%%%%%%%%%%%%%%%%%%%%%%%%%%%%%%%%%

As a second application of the joint track function formalism, we can consider a different detector. On hadronic states, the operators $\mathcal{E}_+$ and $\mathcal{E}_-$ are separately well defined. However, we do not believe they are well defined in a generic field theory, or at least they are not natural. However, if we have a conserved U$(1)$ charge (such as the electromagnetic charge), we can consider the charge operator
\begin{align}
    \cQ(\vec n_1) = 
    \lim_{r\rightarrow \infty} \int \mathrm{d}t \,r^2 n_1^i \,
    J_{i}(t,r\vec{n}_1)\,.
    \label{eq:def}
\end{align}
This detector was discussed in \cite{Hofman:2008ar}, and has been studied in detail in \cite{Chicherin:2020azt}. Correlations of charge flux are not infrared and collinear safe in QCD. While neither $\cQ(\vec n_1) $ or $\mathcal{E}_{\pm}(\vec n_1)$ can be computed in perturbation theory, a major distinction between the two of them is that the $\cQ(\vec n_1)$ is not \emph{soft} safe, due to the lack of energy weighting. This means that divergences associated with correlation functions of these operators are not process independent. This is in distinction with collinear singularities, which are process independent, and can therefore be associated with the detector. From the CFT perspective, the detector $\cQ(\vec n_1)$ has $J=1$ \cite{Kologlu:2019mfz}, and therefore we do expect to be able to make sense of it. In general, it is both phenomenologically and theoretically interesting to understand the space of field theoretically well-defined detectors as illustrated in the Fig.~\ref{fig:detectorspace}, which has primarily been studied in the context of conformal field theories (CFTs) (see ~\cite{Hofman:2008ar,Belitsky:2013xxa,Belitsky:2013bja,Kravchuk:2018htv,Caron-Huot:2022eqs}).

Instead of considering the operator $\cQ(\vec n_1)$, we can consider a detector formed from the product of the $\cQ$ and $\cE$ detectors, defining an detector which we denote $\cE_{\cQ}(\vec n_1)$. By this operator, we mean that when acting on a free particle state, it returns the energy multiplied by the charge
\begin{align}
\cE_{\cQ}(\vec n_1)|k\rangle=E_k Q_k \delta(\vec n_1-\vec \hat k) |k \rangle.
\end{align}
 Since this detector involves a product of operators, it must be renormalized. This is similar to products of $\cE$ operators considered in \cite{Caron-Huot:2022eqs}. We will show that this operator can be sensibly renormalized in perturbation theory. Although we will not consider it here, it would be interesting to study the more formal properties of this operator. We believe that it should exist non-perturbatively. Unlike most detectors studied so far in the literature, this detector is $C$-odd, giving a different result on particles and anti-particles, and thus should live on a different Regge trajectory than the $\cE$ detector. Due to the interesting phenomenological applications of such operators as we will show, we believe that developing a more formal understanding of their properties is worthwhile.

From a phenomenological perspective, this operator is quite interesting due to the fact that the result that it returns when acting on a particle state can be either positive, or negative. We will see that this leads to a qualitatively different scaling behavior in multi-point correlation functions, as compared to either the standard $\cE$ operators, or the $\cE_{\pm}$ operators.

\begin{figure}
  \begin{center}
  \subfloat[]{
  \includegraphics[scale=0.26]{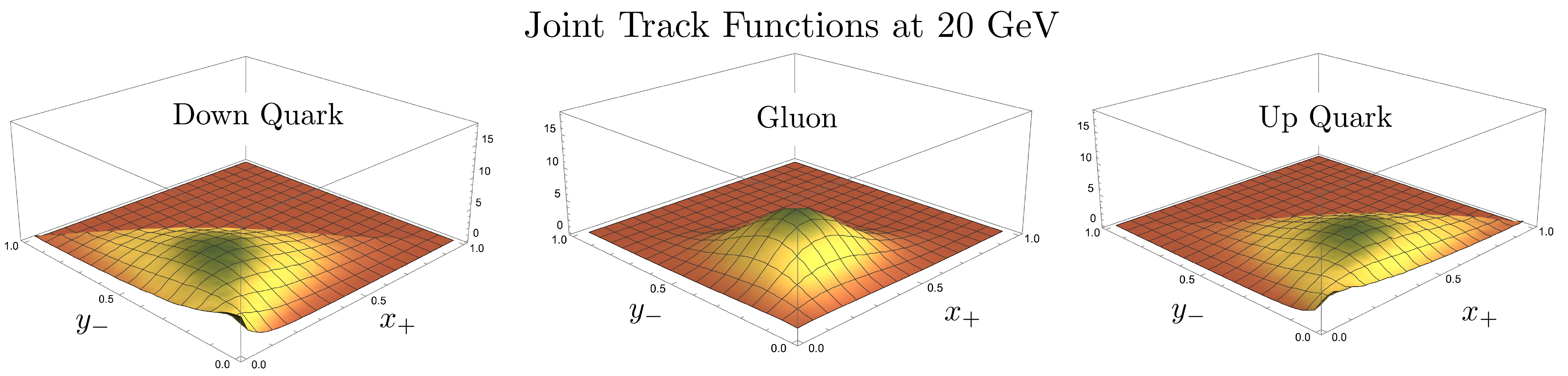}
  \label{fig:ratioNto2_3pt_a}
  }\\
    \subfloat[]{
  \includegraphics[scale=0.26]{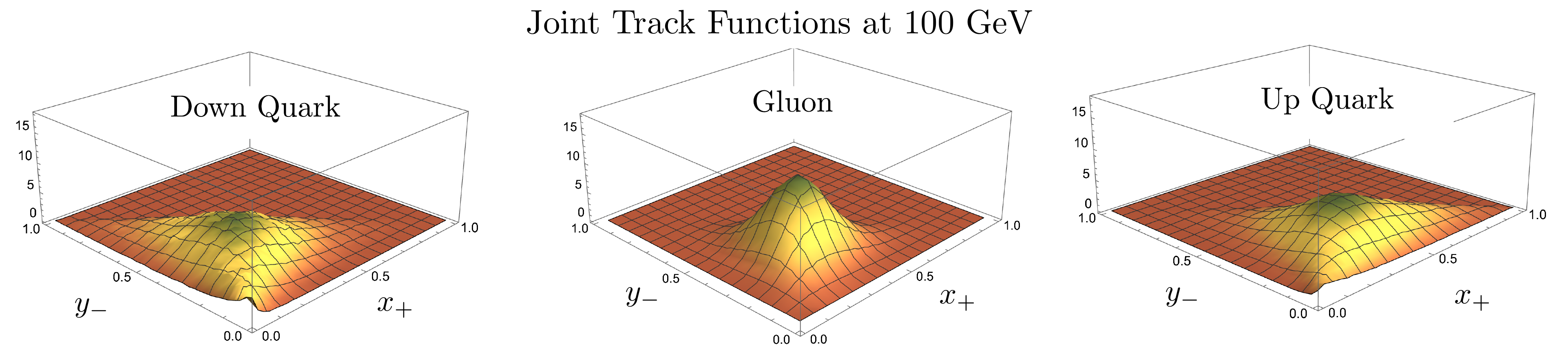}
  \label{fig:ratioNto2_3pt_a}
  }\\
    \subfloat[]{
  \includegraphics[scale=0.26]{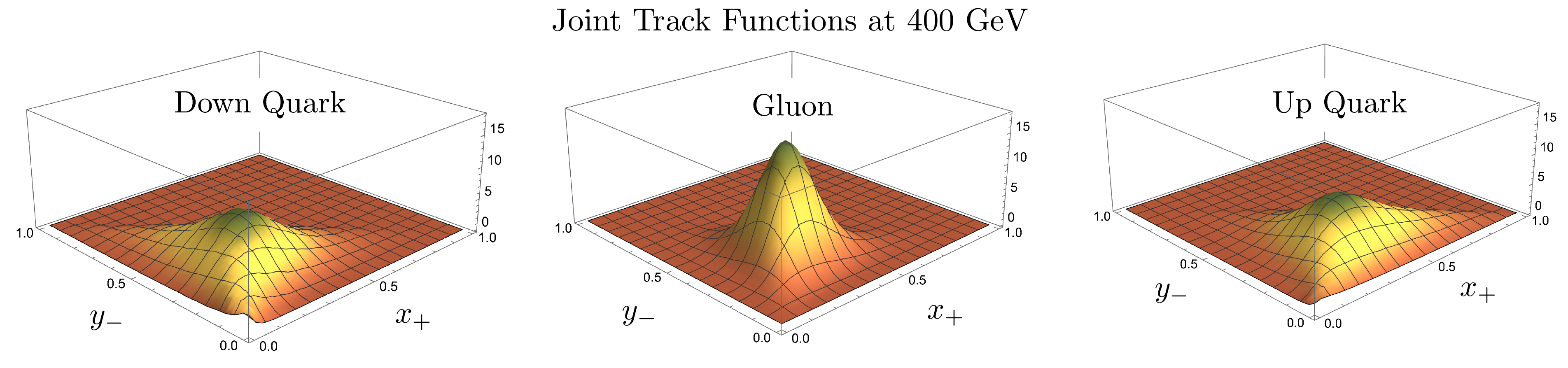}
  \label{fig:ratioNto2_3pt_a}
  }
  \end{center}
  \caption{The evolution of the joint track functions, $T_i^{+-}(x_+,y_-;\mu)$, as extracted from the Pythia parton shower. We clearly observe that up type quarks deposit more of their energy into positively charged hadrons, down type quarks deposit more energy into negatively charged hadrons, and gluons deposit their energy symmetrically. As the scale is increased, the covariance of the distribution decreases, in accordance with the predictions of the renormalization group equations. }
  \label{fig:joint_track_pythia}
  \end{figure}

A natural object to study are $N$-point projected $\mathcal{E}_Q$-correlators. We can  express these in terms of the charge correlators as
\bea
\langle\underbrace{\mathcal{E}_Q\mathcal{E}_Q ... \mathcal{E}_Q}_{N \text{ times}} \rangle = \sum_{k=0}^N \binom{N}{k}(-1)^k \langle \underbrace{\mathcal{E}_+ \cdots \mathcal{E}_+}_{N-k \text{ times}} \underbrace{\mathcal{E}_- \cdots \mathcal{E}_-}_{k \text{ times}} \rangle\,.
\eea
That is, the $N$-point $\mathcal{E}_Q$-correlators probes the asymmetry between the $+$ and $-$ charge distribution through the $N$-th moment of the 1-dimensional distribution with $Z=X-Y$ as its random variable. This is $T^{\rm asy}(N;\mu)$ defined above. We are therefore also able to factorize them in an identical manner to the multipoint projected correlators of the $\cE_+$ and $\cE_-$ operators. The only aspect of the factorization theorem that changes, is that this observable is sensitive to the asymmetry of the joint track function, as opposed to the central moments. At LL order, we can therefore write the jet function as
\begin{align}
\label{eq:chargeprodLL}
&\vec{J}^{\cE_\cQ}_{\rm LL}\left(\ln \frac{p_T^2 R_L^2}{\mu^2}, \mu\right)=
\vec{T}^{\mathrm{asy}}(2;p_T R_L)
\cdot \left(\frac{\alpha_s(p_T R_L)}{\alpha_s(\mu)}\right)^{-\frac{\hat{\gamma}^{(0)}(3)}{\beta_0}}\,,
\end{align}
where the vector notation is similar to the one defined above in Eq.~\eqref{eq:Jbdry}.
We find it interesting how the different moments of the joint track functions appear in the factorization theorems for different physical observables. This suggests that the joint track functions provide the appropriate generalization of the track functions that can be used to describe a broad range of observables.

\begin{figure}
    \subfloat[]{
\includegraphics[scale=0.25]{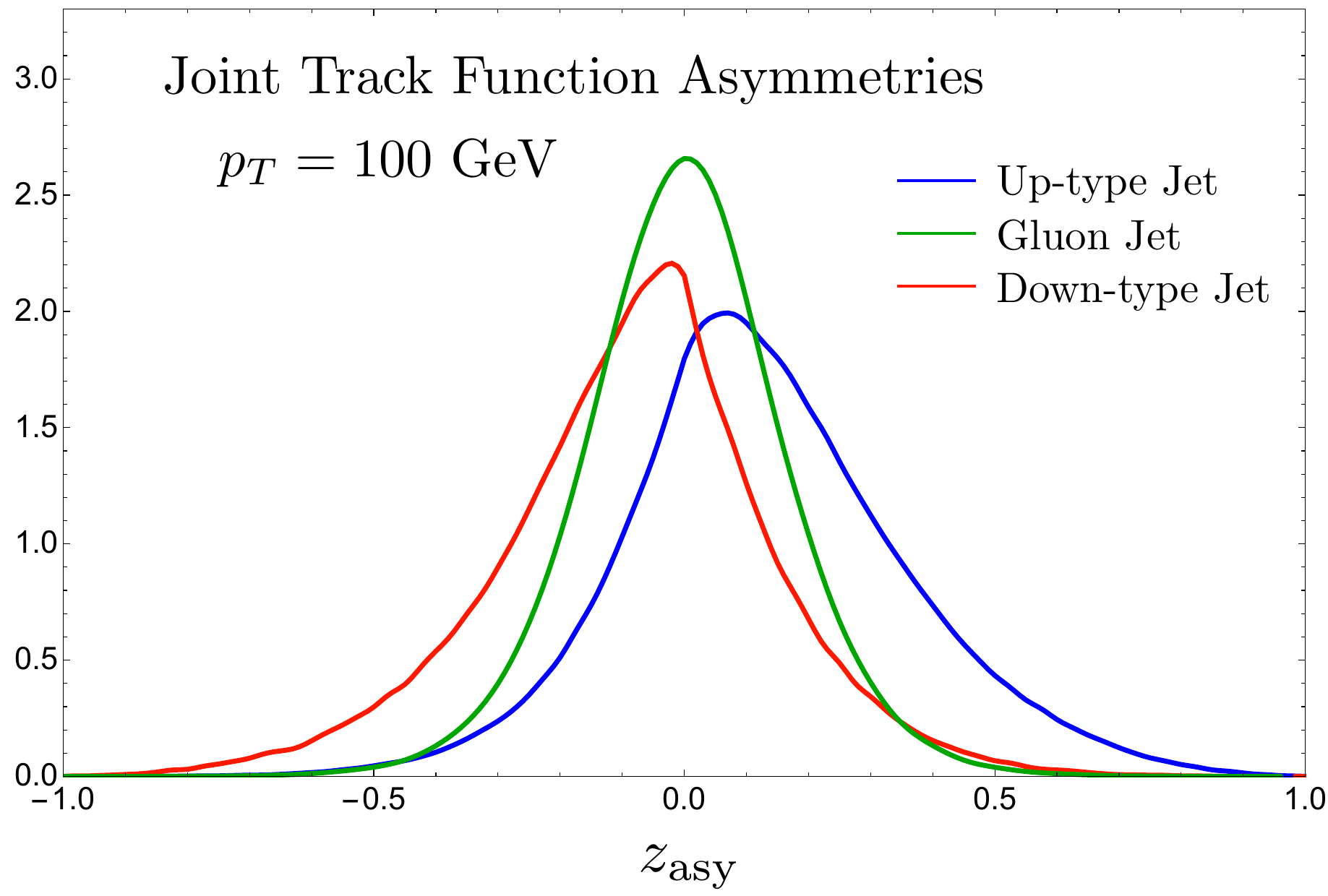} 
  \label{fig:asym_a}
  }
      \subfloat[]{
\includegraphics[scale=0.254]{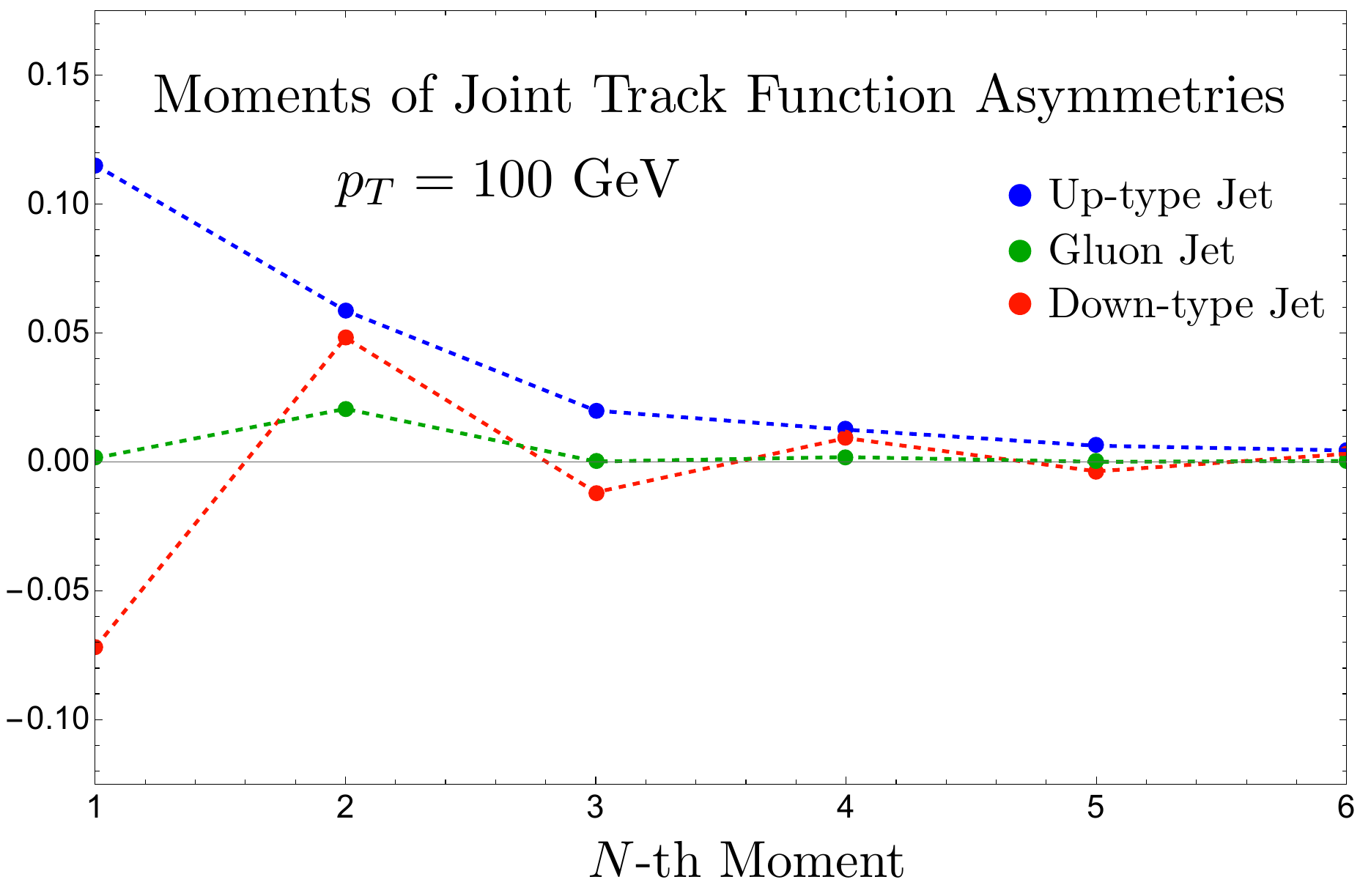} 
  \label{fig:asym_b}
  }
\caption{(a) The asymmetry track function inside quark and gluon jets at $p_T=100$ GeV. (b) Moments of the quark and gluon asymmetry track functions $T_i^{\mathrm{asy}}(z)$ at $p_T=100$ GeV. Due to the asymmetry track function's close resemblance to an even function, as illustrated in Fig.~\ref{fig:asym}, the odd moments exhibit suppression compared to the even moments.}
\label{fig:asym}
\end{figure}

\subsection{Generic Energy Correlation Functions}
Although our focus is mainly on the case involving $+$ and $-$ charges for phenomenological studies, the framework presented above readily facilitates the extension of this formalism for the study of generic energy correlation functions involving multiple quantum numbers.

Considering $k$ distinct quantum numbers, the most general energy correlators, which examine the flux of these quantum numbers, are expressed as
\bea
\langle \underbrace{\mathcal{E}_{R_1}\cdots \mathcal{E}_{R_1}}_{N_1 \text{ times}} \underbrace{\mathcal{E}_{R_2} \cdots \mathcal{E}_{R_2}}_{N_2 \text{ times}}\cdots  \underbrace{\mathcal{E}_{R_k} \cdots \mathcal{E}_{R_k}}_{N_k \text{ times}}\rangle\,.
\eea
The cumulant of the projected $(N_1,N_2,\cdots,N_k)$-point charge correlators $\Sigma_{R_1\, R_2\cdots R_k}^{[N_1,N_2,\cdots,N_k]}\left(R_L, p_T^2, \mu\right)$ then factorizes as 
\begin{align}
& \Sigma_{R_1\, R_2\cdots R_k}^{[N_1,N_2,\cdots,N_k]}\left(R_L, p_T^2, \mu\right)=\int_0^1 d x~ x^{N_1+N_2+\cdots+N_k} \vec{J}_{R_1\, R_2\cdots R_k}^{[N_1,N_2,\cdots,N_k]}\left(R_L, x,  \mu\right) \cdot \vec{H}\left(x, p_T^2, \mu\right)\,.
\end{align}

Then by renormalization group consistency, the general energy correlator jet function obeys the renormalization group evolution equation
\bea
\frac{d \vec{J}_{R_1\, R_2\cdots R_k}^{[N_1,N_2,\cdots,N_k]}\left(\ln \frac{p_T^2 R_L^2}{\mu^2}\right)}{d \ln \mu^2}=\int_0^1 dy \, y^{N_1+N_2+\cdots+N_k}  \vec{J}_{R_1\, R_2\cdots R_k}^{[N_1,N_2,\cdots,N_k]}\left(\ln \frac{y^2 p_T^2 R_L^2}{\mu^2}\right) \cdot \widehat{P}(y)\,,
\eea
where $\widehat{P}(y)$ is the singlet timelike DGLAP splitting kernel matrix. The jet function $\vec{J}$ (and similarly for the hard function) is a vector in flavor space, analogous to Eq.~\eqref{eq:Jbdry}, and is expressed in terms of the moments of the most generic joint track function as given in Eq.~\eqref{eq:jtmom}. For instance, the boundary condition for the jet function at tree level will involve
\bea
T^{R_1\,R_2 \cdots R_k}_i(N_1, N_2,\cdots,N_k;p_T R_L)\,.
\eea
We posit that the most generic energy correlation functions formalism presented herein will open exciting opportunities and phenomenological applications to probe the intricate correlations between the most generic flux of quantum numbers with precision.

%%%%%%%%%%%%%%%%%%%%%%
\section{Numerical Results for Joint Track Functions}\label{sec:numeric}
%%%%%%%%%%%%%%%%%%%%%%  

Although the focus of this paper is on setting up the formalism to describe joint track functions, to make them more concrete, and to illustrate that they can be used to compute observables, it is interesting to extract them and apply them to the calculation of different energy correlator observables. Ideally this would be done using data, but here we consider the simple case of extracting them from the parton shower Monte Carlo parton shower Pythia \cite{Sjostrand:2014zea,Sjostrand:2007gs}. Pythia  implements a string model of hadronization \cite{Andersson:1983ia}. It would be particularly interesting to study how the joint track function depends on the details of the hadronization model, for example, by comparing the string model \cite{Andersson:1983ia} with the HERWIG \cite{Bahr:2008pv,Bahr:2008tf,Bellm:2019zci} implementation of the cluster model \cite{Marchesini:1991ch}. Since the joint track function is sensitive to correlations in the hadronization process, we expect it to be more sensitive to the details of the dynamics of the hadronization process. However, detailed comparisons with different hadronization models are beyond the scope of this paper, and we leave them to future work.

     \begin{figure}
 \begin{center}
     \subfloat[]{
\includegraphics[scale=0.27]{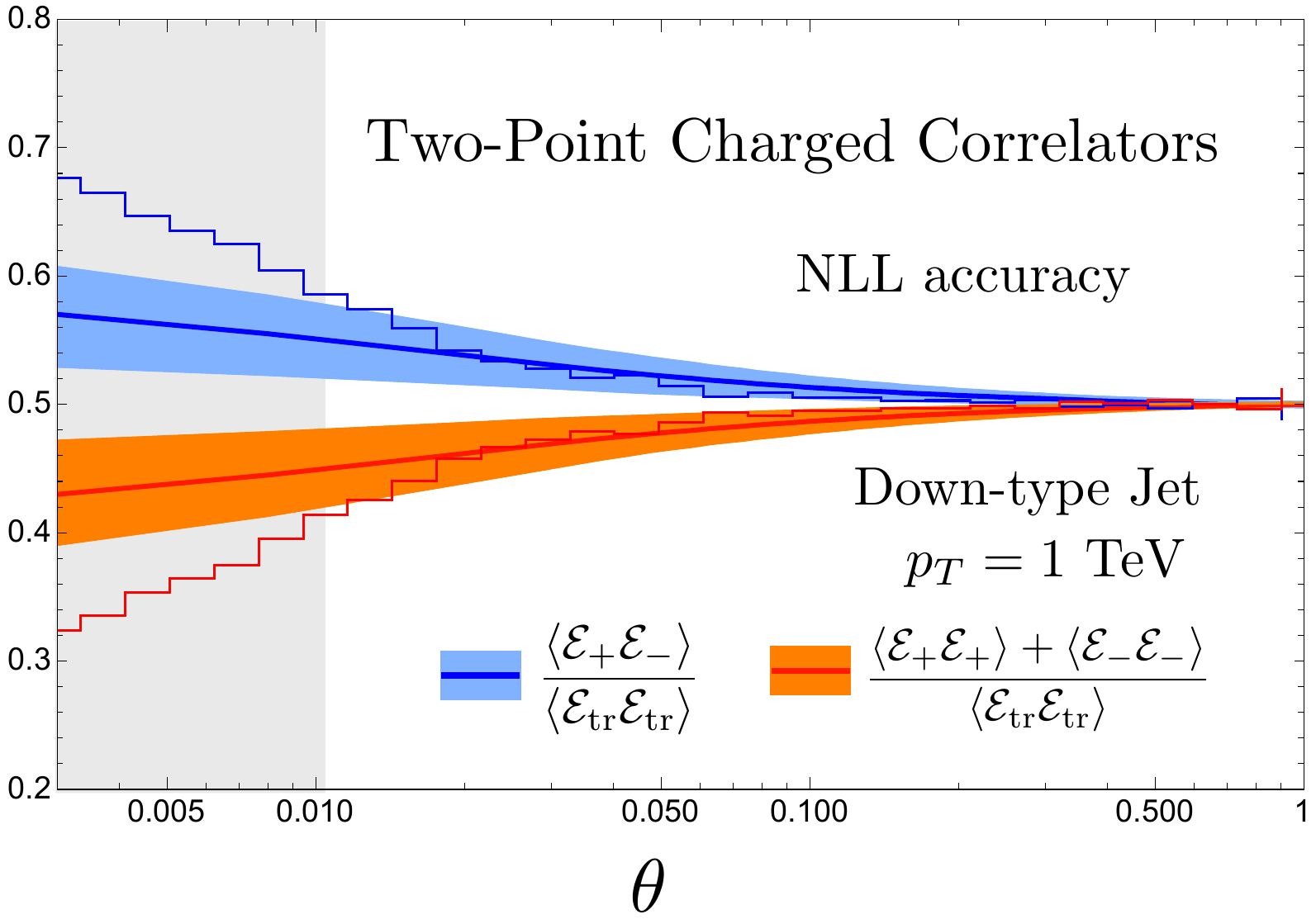} 
\label{fig:pm_plot}
}
     \subfloat[]{
 \includegraphics[scale=0.267]{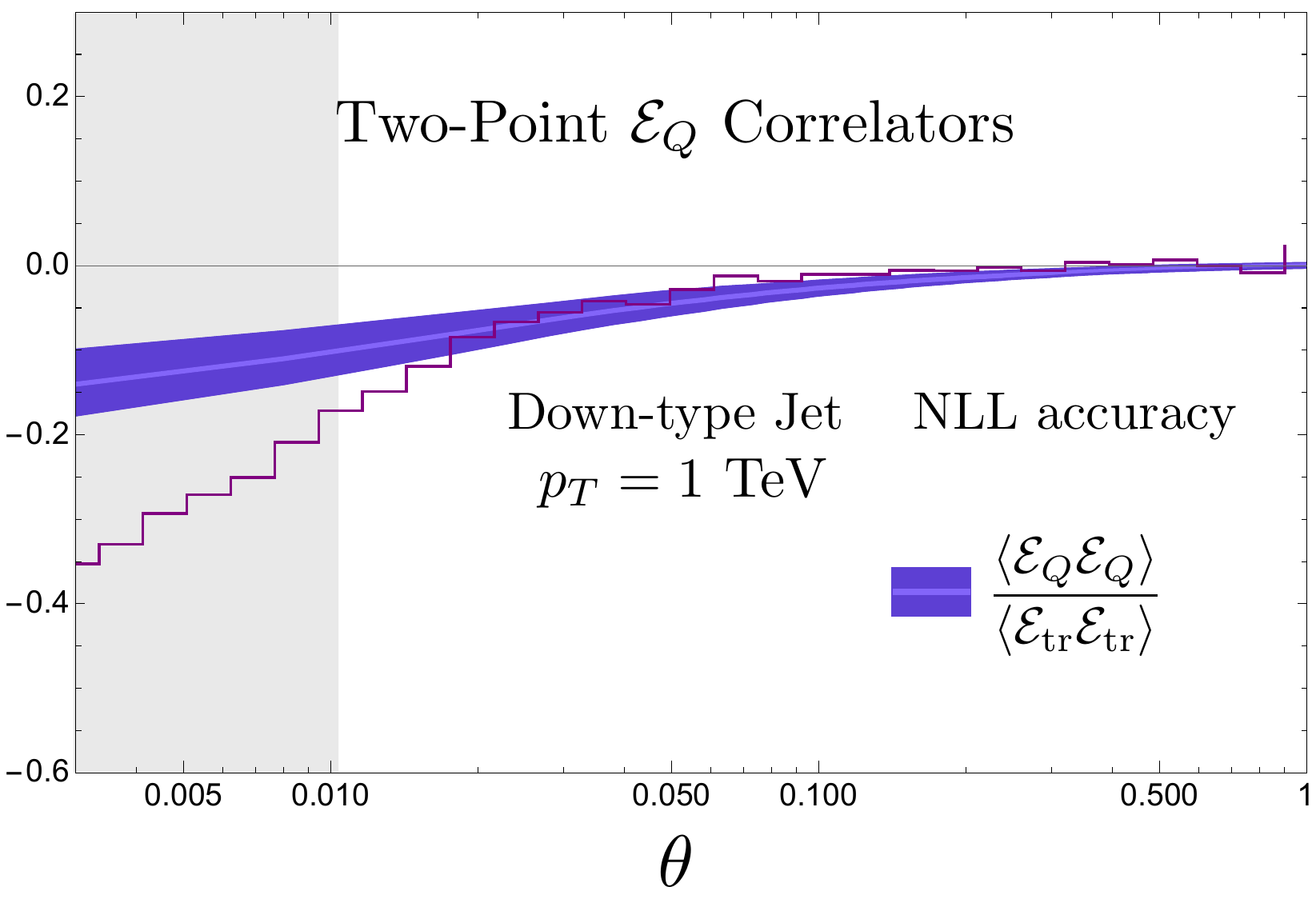}   
\label{fig:EQ_2point}
}\\
     \subfloat[]{
\includegraphics[scale=0.27]{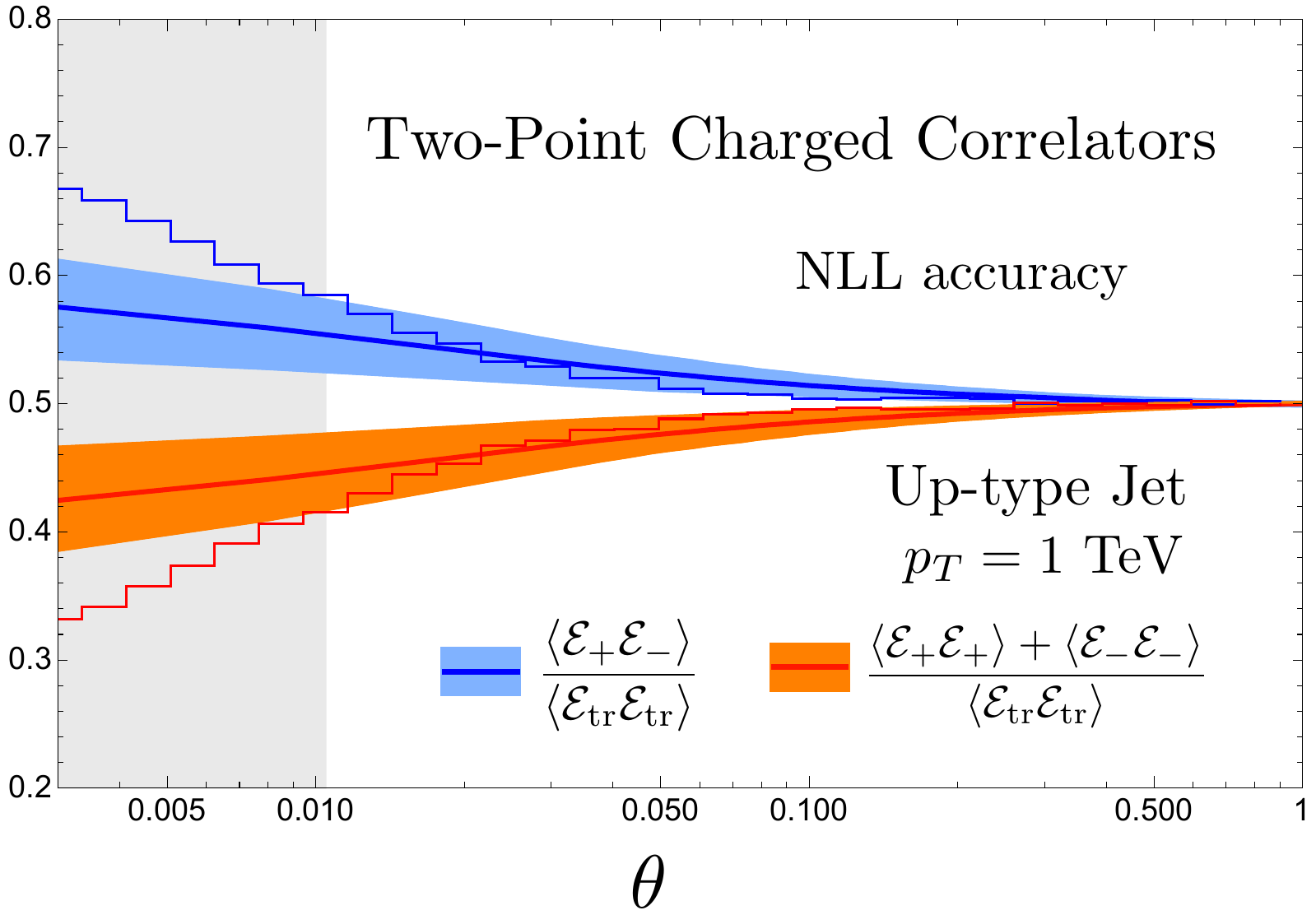} 
\label{fig:pm_plot_up}
}
     \subfloat[]{
 \includegraphics[scale=0.267]{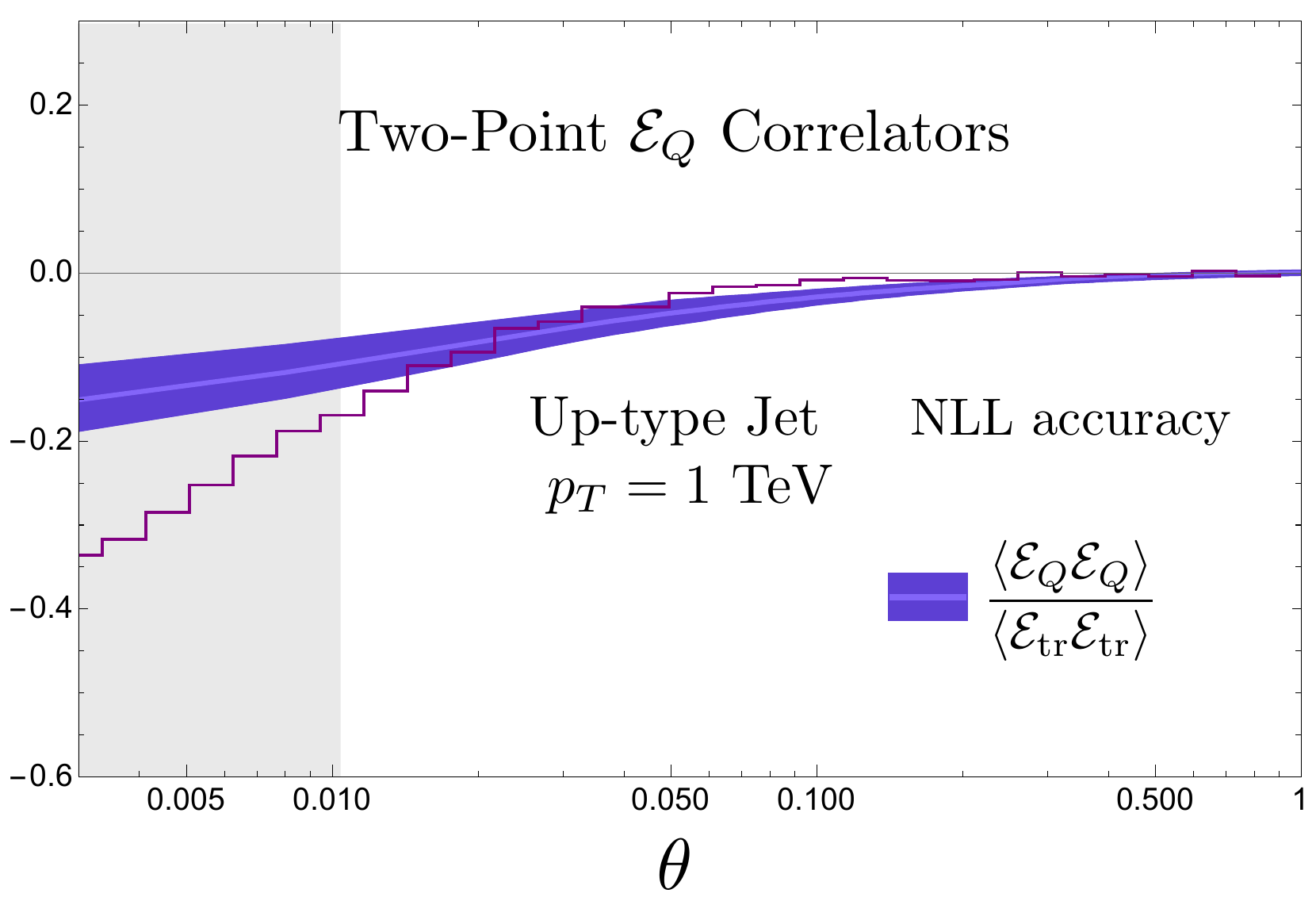}   
\label{fig:EQ_2point_up}
}\\
     \subfloat[]{
\includegraphics[scale=0.27]{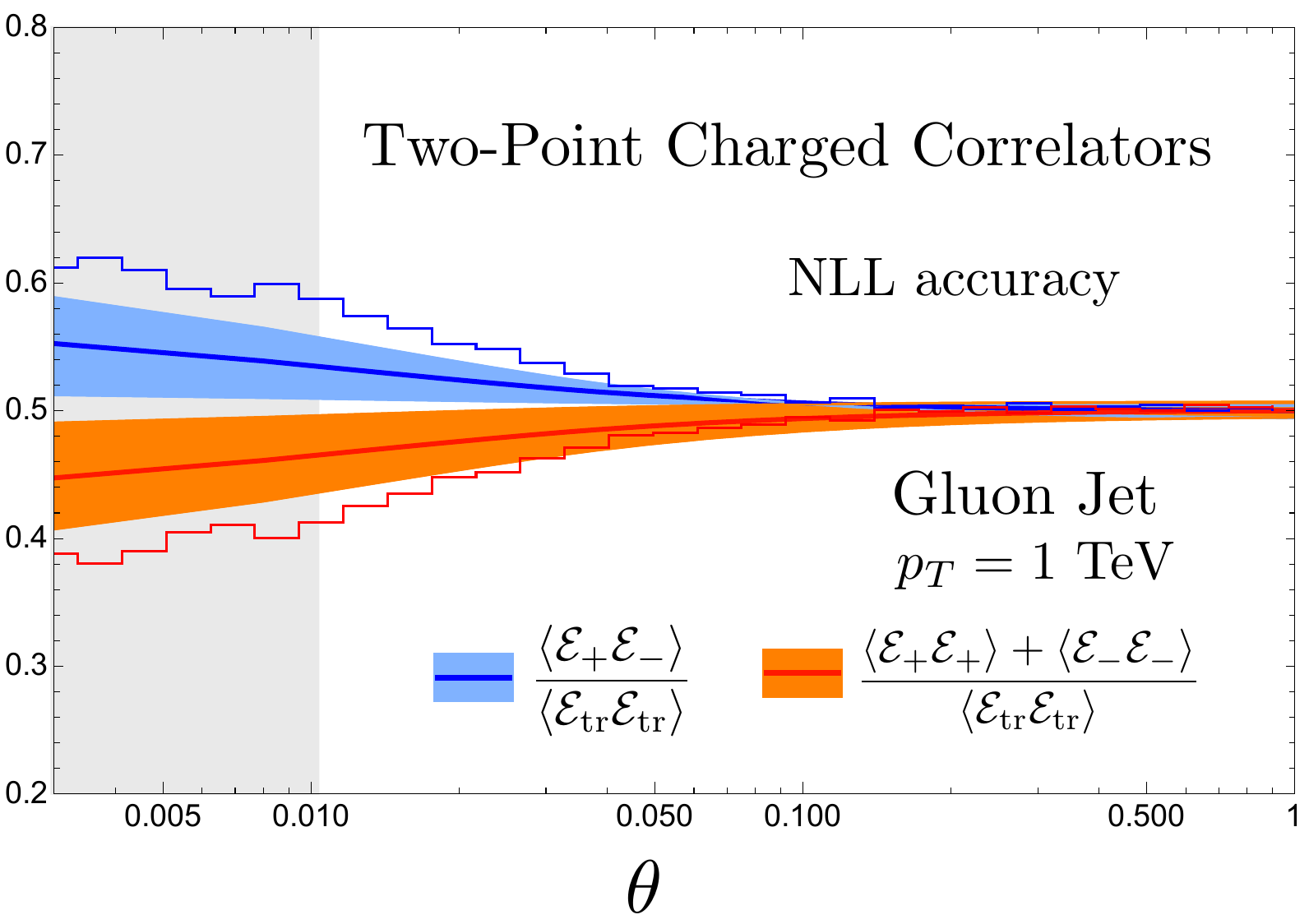} 
\label{fig:pm_plot_gluon}
}
     \subfloat[]{
 \includegraphics[scale=0.267]{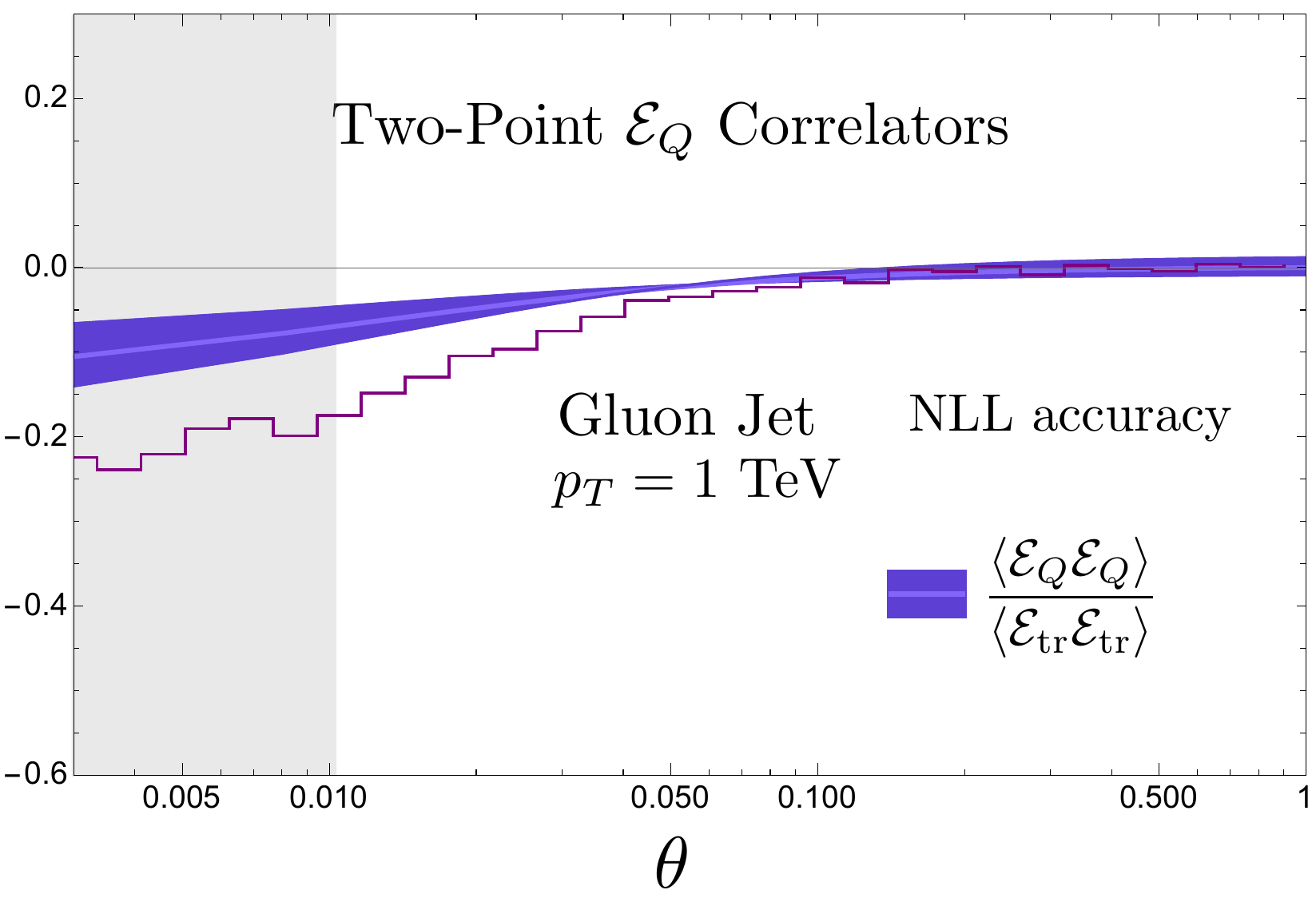}   
\label{fig:EQ_2point_gluon}
}
\end{center}
\vspace{-0.45cm}
\caption{Left column: A comparison of the $\langle \cE_+ \cE_-\rangle $ and $\langle \cE_+ \cE_+\rangle$ correlators for down type quark jets (a), up type quark jets (c), and gluons (e). Results from our NLL calculations are shown in solid, and results from Pythia are shown in histograms. Enhanced correlations are observed for opposite signed hadrons as compared to same sign hadrons. Right Column: The two point correlator $\langle \cE_\cQ \cE_\cQ \rangle$ for down type quark jets (b), up type quark jets (d) and gluons (f). For both columns, the transition to the non-perturbative regime is shown in gray. }
\label{fig:2point_plots}
\end{figure} 

In \Fig{fig:joint_track_pythia} we show the joint track function $T^{+-}(x_+,y_-)$ as extracted from Pythia for up type quarks, down type quarks, and gluons. The distributions are shown at three different energies, $20$ GeV, $100$ GeV, and $400$ GeV, to show the evolution as a function of scale. The results are quite remarkable. First, we see clearly that the gluon joint track function is symmetric as a function of $x_+,y_-$, as is required by charge conjugation invariance. On the other hand, we see that down type quarks deposit a larger fraction of energy into negatively charged hadrons, while up type quarks deposit a large fraction of energy into positively charged hadrons. As we evolve the distributions to higher energy values, we clearly see that the distributions become more peaked, moving towards the limit of zero covariance in the UV, as predicted by the renormalization group evolution. This is particularly prominent for the gluon, which evolves the fastest due to its larger anomalous dimension. These joint track functions provide a fascinating insight into correlations in the hadronization process.

In Fig.~\ref{fig:asym_a}, we project the joint track functions onto the asymmetry track distribution. We consider the case of  $\mu=100$ GeV as an example. For gluons this function is even, while up type and down type quarks exhibit a slight asymmetry. Because of this, there is a large hierarchy between the odd and even moments of the asymmetry distribution, as seen in Fig.\ref{fig:asym_b}. These lead to qualitatively different behavior for $N$-point correlators of the $\cE_\cQ$ operators depending on whether $N$ is even or odd.

It should be possible to measure the joint track function at the LHC by measuring the fraction of positively and negatively charged hadrons in a high energy jet. We are unaware of measurements of this nature being performed at previous experiments. We believe that this would be a particularly impactful measurement, since it would enable the calculation of a much broader class of multipoint correlation functions involving electromagnetic charge. 

    \begin{figure}
\begin{center}
     \subfloat[]{
\includegraphics[scale=0.28]{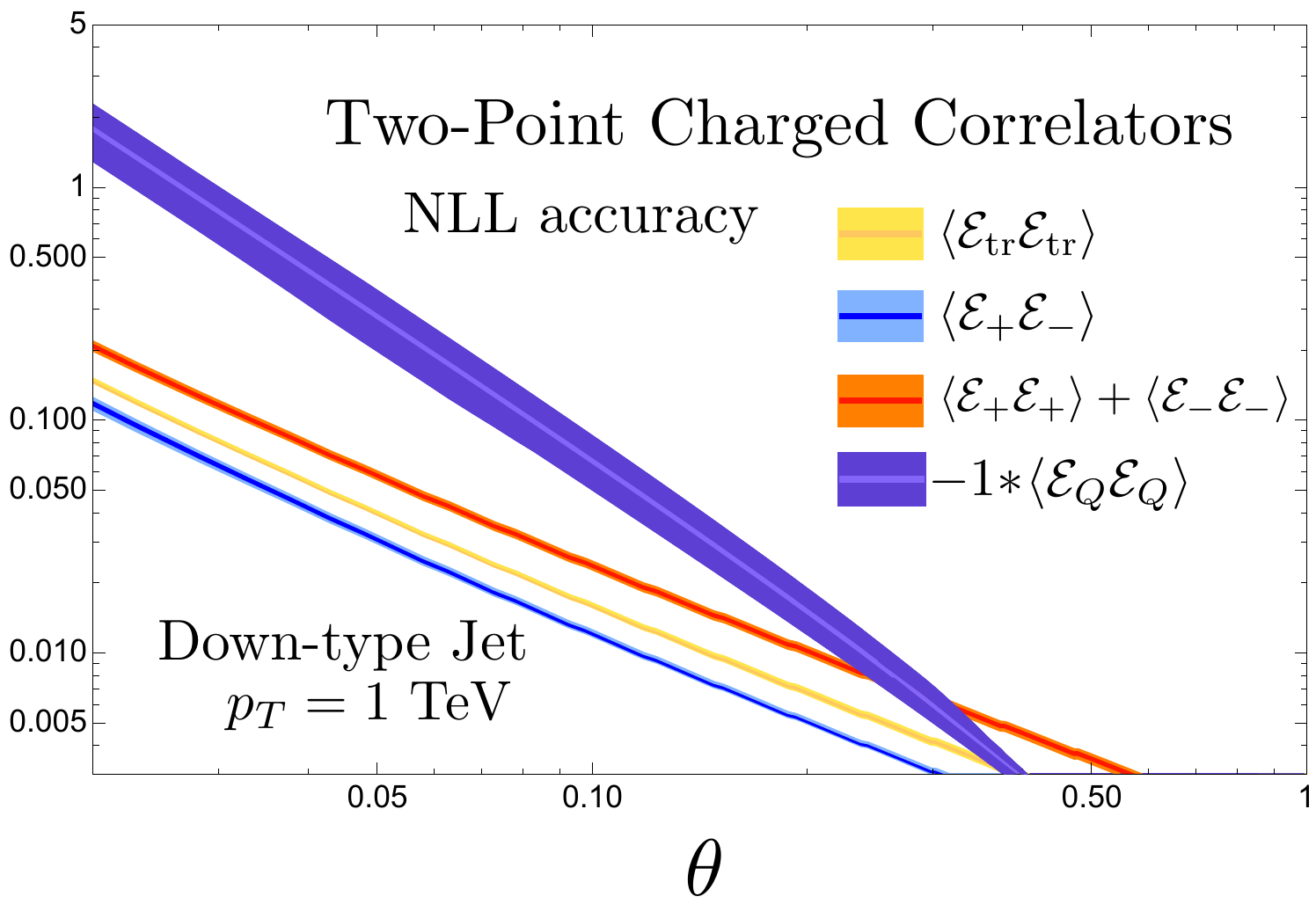} 
}\\
     \subfloat[]{
\includegraphics[scale=0.28]{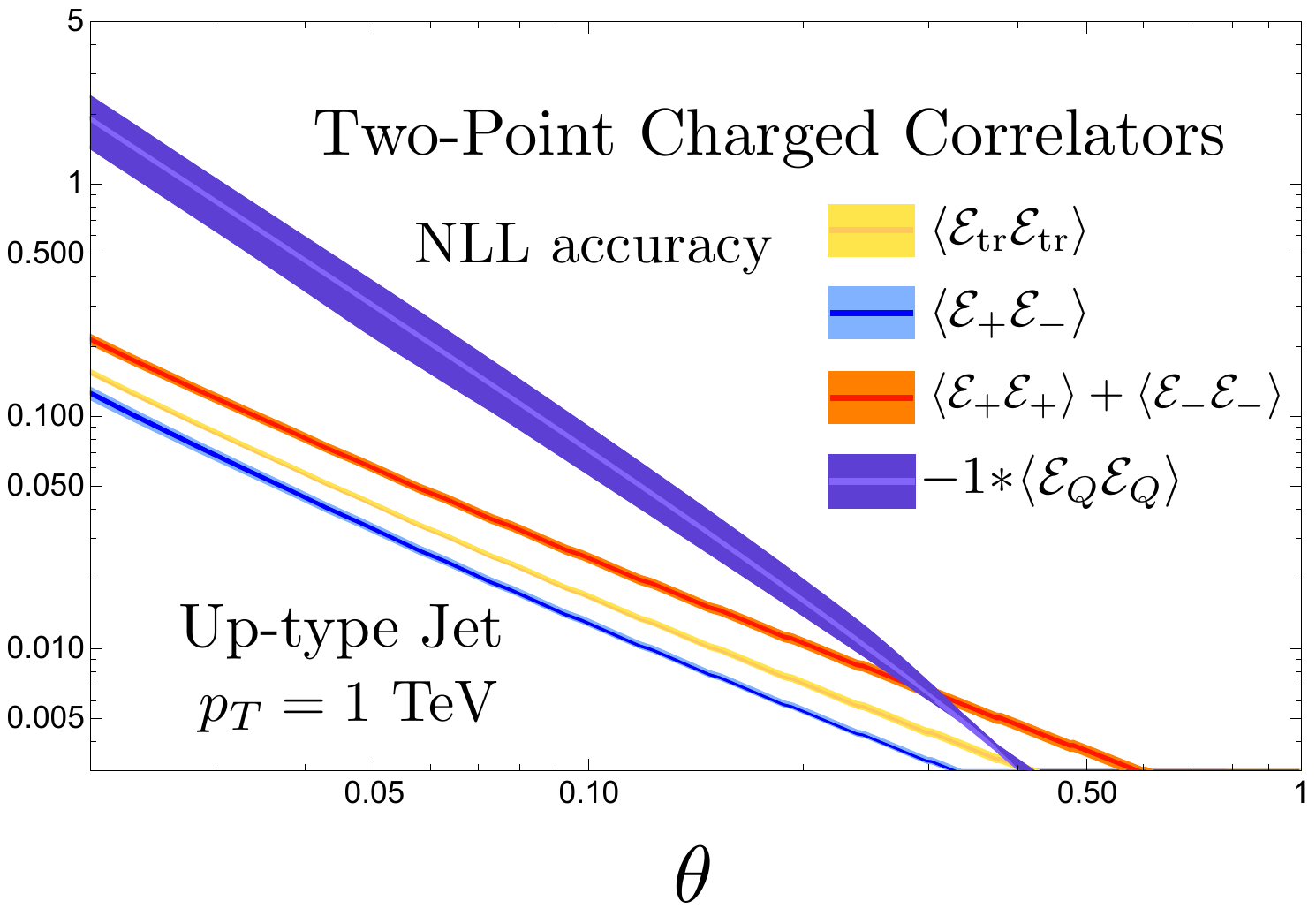} 
}\\
     \subfloat[]{
\includegraphics[scale=0.28]{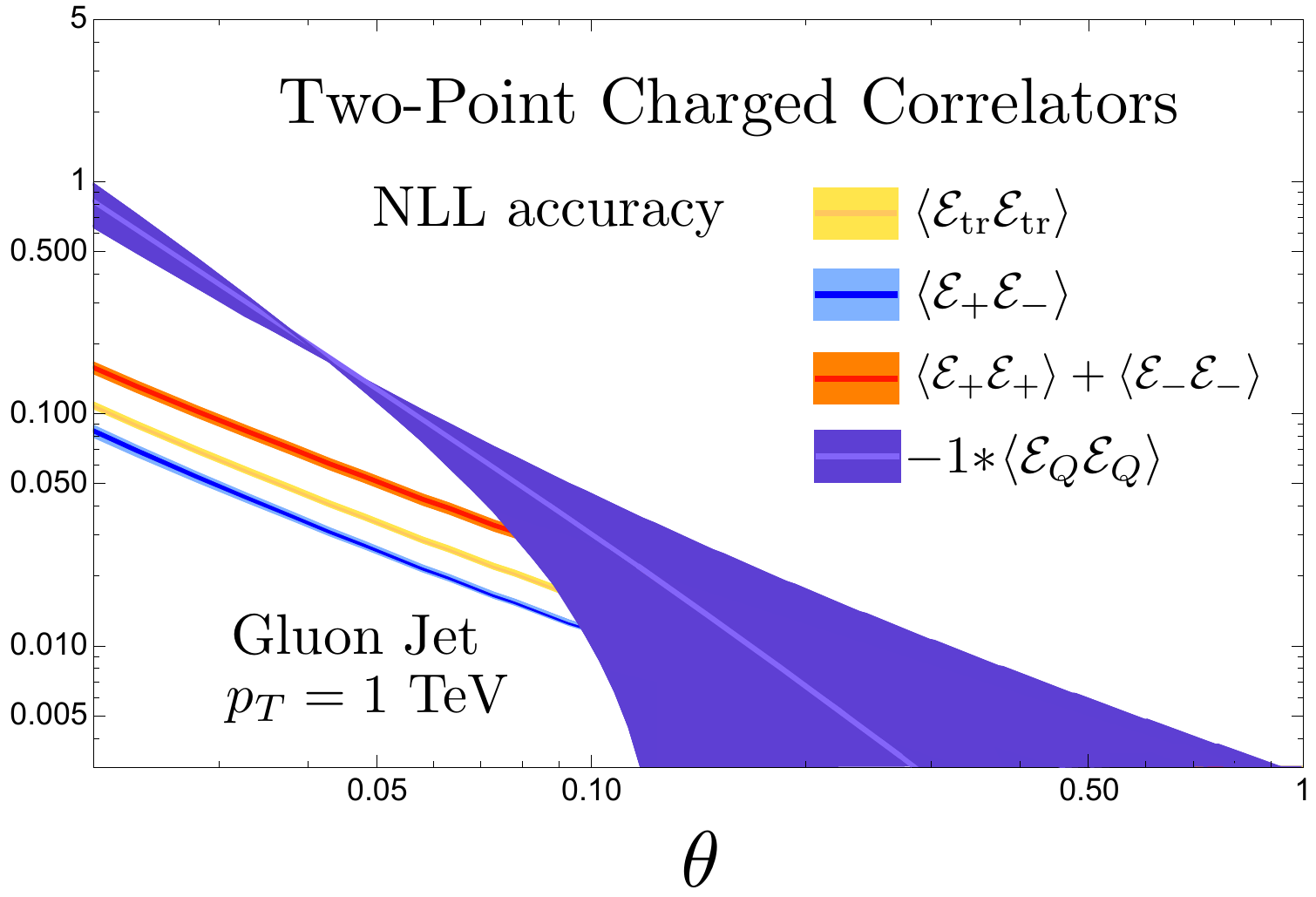} 
}
\end{center}
\caption{A logarithmic plot of the two-point correlators formed from the $\cE_+$, $\cE_-$ and $\cQ$ correlators for down type quarks (a), up type quarks (b) and gluons (c). Due to cancellations, the two point correlator of $\cQ$ operators exhibits a power law of  $d\sigma/d\theta \sim \theta^{-2}$ up to logarithmic corrections, as compared to the more standard $d\sigma/d\theta \sim \theta^{-1}$. This is correctly predicted by the joint track function formalism.}
\label{fig:EQ_log}
\end{figure}

%%%%%%%%%%%%%%%%%%%%%%
\section{Scaling Behavior of $\cE_{\pm}$ and $\cE_{\cQ}$ Correlators}\label{sec:numeric_correlators}
%%%%%%%%%%%%%%%%%%%%%%  

Having extracted the joint track functions, we can now apply them to the calculation of a variety of multi-point energy correlators that incorporate charge. As emphasized above, these distributions do not involve the complete distribution, but rather only require certain moments. We will see that these different correlators provide fascinating insights into the correlations of charge within jets. As the primary focus of our paper is extending the formalism to generic correlators involving multiple quantum numbers, we presented our equations at LL accuracy for clarity. However, this can be easily extended to NLL accuracy in a similar fashion to~\cite{Chen:2020vvp} and the results presented in this section are computed to collinear NLL accuracy and incorporates the leading non-perturbative power corrections.

 We begin by considering charged two-point correlators. In \Fig{fig:pm_plot},  \Fig{fig:pm_plot_up} and \Fig{fig:pm_plot_gluon} we show the two different correlators $\langle \cE_+ \cE_-\rangle$ and $\langle \cE_+ \cE_+\rangle$ for up-type and down-type quark jets, as well as gluon jets. In all cases we observe enhanced correlations at small angles for opposite signed hadrons as compared to like sign hadrons. Note that these correlations are generated purely from the strong interactions, and are not electromagnetic in nature. We compare our result with Pythia (shown as a histogram). Overall, good agreement is observed, with deviations becoming larger as the distribution goes deeper into the non-perturbative regime. We note that the two-point correlation is largely independent of the charge of the quark initiating the jet. This information is carried more strongly by the one-point functions, $\langle \cE_- \rangle$ and $\langle \cE_+ \rangle$. The two-point functions capture the fluctuations about this.

 In \Fig{fig:EQ_2point}, \Fig{fig:EQ_2point_up} and \Fig{fig:EQ_2point_gluon} we show the two-point correlator $\langle \cE_\cQ \cE_\cQ \rangle$ again on up-type and down-type quark jets, and gluon jets. We compare to Pythia shown as a histogram. Again, we see overall good agreement with deviations becoming larger in the non-perturbative regime. These results show that correlators involving electromagnetic charge exhibit clean scaling behavior, just like standard energy correlators, which can be computed using the joint track functions. This relies crucially on the form of the detectors, namely that they have powers of the energy, and illustrates that the correlator based approach is the correct way of asking about correlations of charge in hadronic collisions. Again, we emphasize that while the one point function $\langle \cE_\cQ  \rangle$ (which is related to average jet charge \cite{Krohn:2012fg,Waalewijn:2012sv}) depends strongly on the charge of the initiating state, the two-point correlations do not. This is due to the fact that a single detector is $C$-odd, and therefore gives an opposite result for e.g. up quark and anti-up quark jets, while the two-point function is $C$-even, and therefore gives the same result. We will see that the overall charge of the state is reflected in the three-point function, which we will study shortly.
 
   \begin{figure}
\begin{center}
     \subfloat[]{
\includegraphics[scale=0.27]{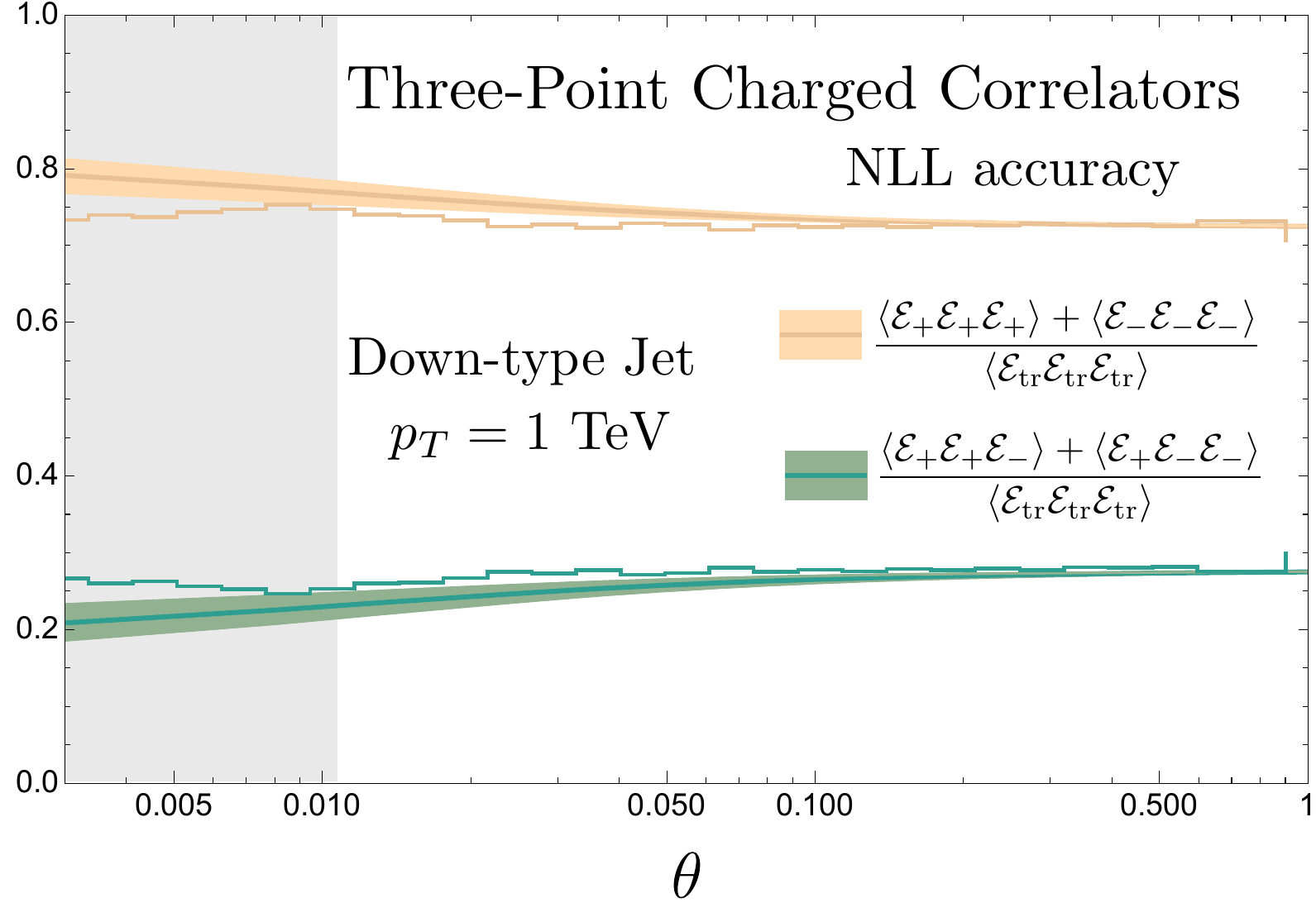} 
}\\
     \subfloat[]{
\includegraphics[scale=0.27]{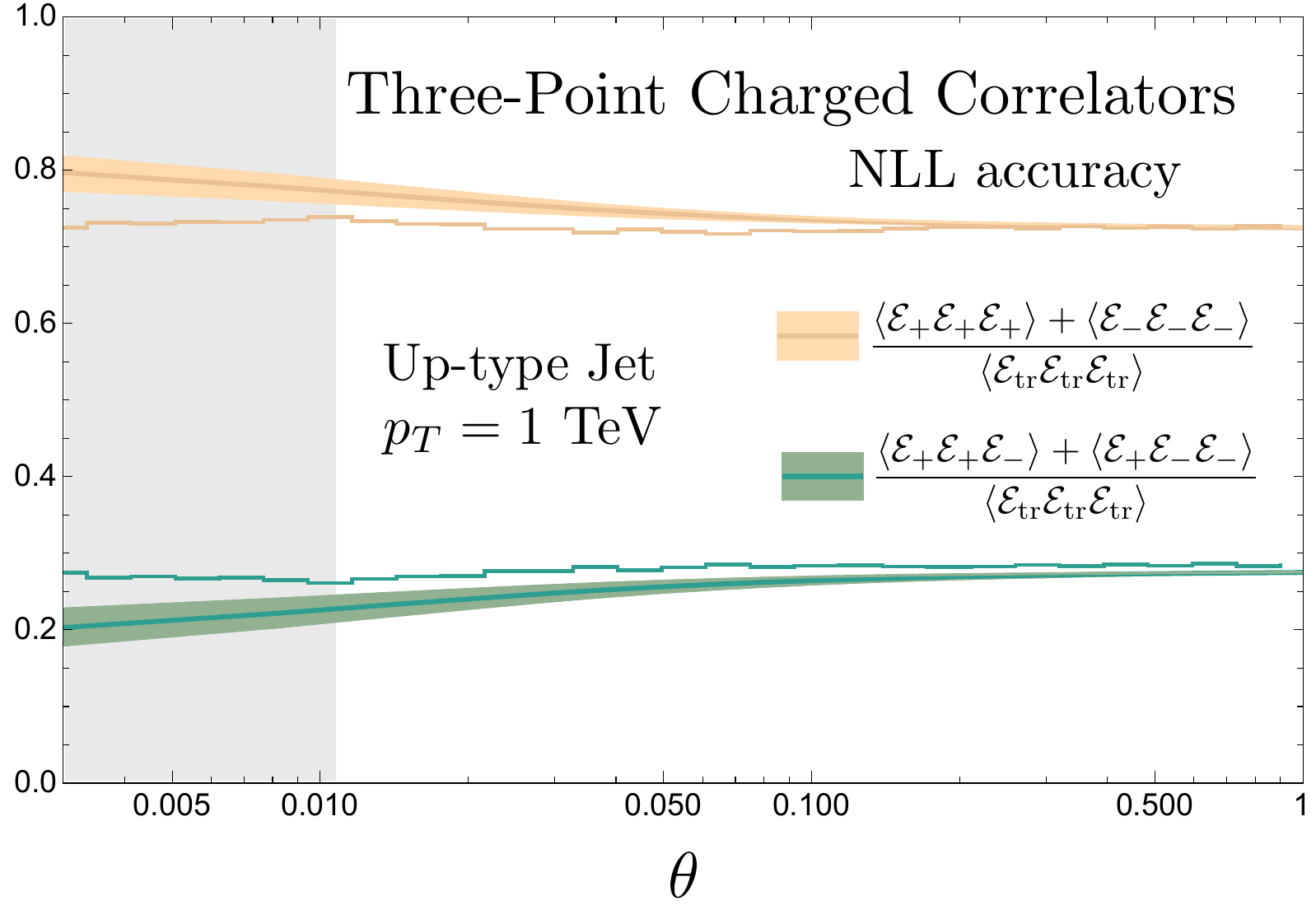} 
}\\
     \subfloat[]{
\includegraphics[scale=0.27]{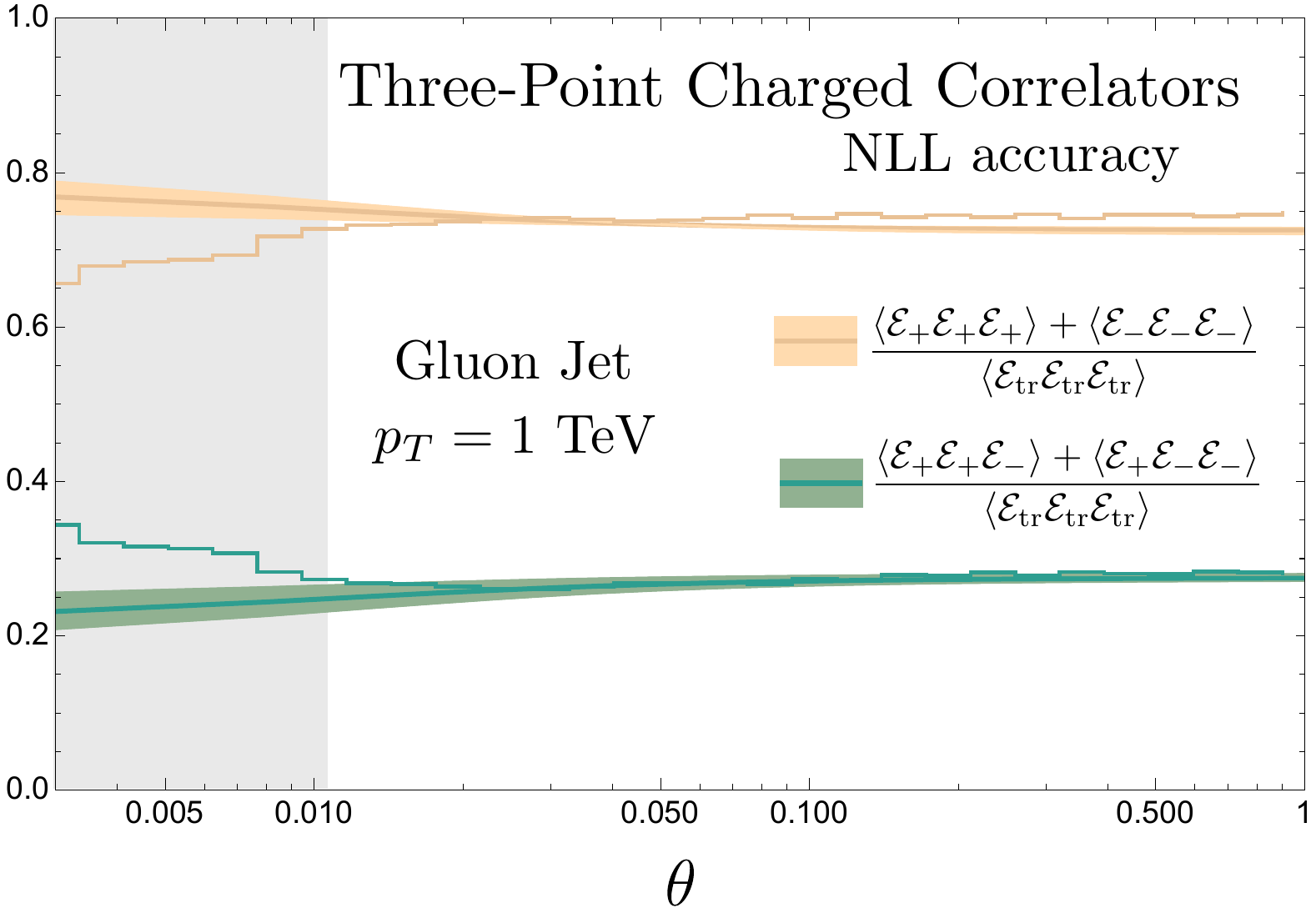} 
}
\end{center}
\vspace{-0.25cm}
\caption{Three point correlators involving the $\cE_+$ and $\cE_-$ detectors for down-type quark jets (a), up-type quark jets (b) and gluon jets (c). Results from our theoretical calculations are shown in solid, and results from Pythia are shown in histograms. The transition to the non-perturbative regime is shown in gray. Slightly different behaviors are observed for the two different combinations of detectors. }
\label{fig:pm_higherpoint}
\end{figure} 

A particularly interesting feature of the $\cE_{\cQ}$ detectors is that they lead to a qualitatively different power law behavior than either the $\cE$ operators or the $\cE_{\pm}$ operators, due to the fact that the $\cE_{\cQ}$ operator can be either positive, or negative, which introduces large cancellations. In particular, while the two-point correlator of energy flow operators scales like $d\sigma/d\theta \sim \theta^{-1}$ up to logarithmic corrections, the two point correlator of $\cE_\cQ$ operators scales like $d\sigma/d\theta \sim \theta^{-2}$ up to logarithmic corrections. This is shown in the log-log plot in \Fig{fig:EQ_log}. We see that the joint track function formalism is able to describe this scaling. It would be interesting to understand how this different scaling law can be directly seen from the lightray OPE perspective.

\begin{figure}
\begin{center}
\includegraphics[scale=0.35]{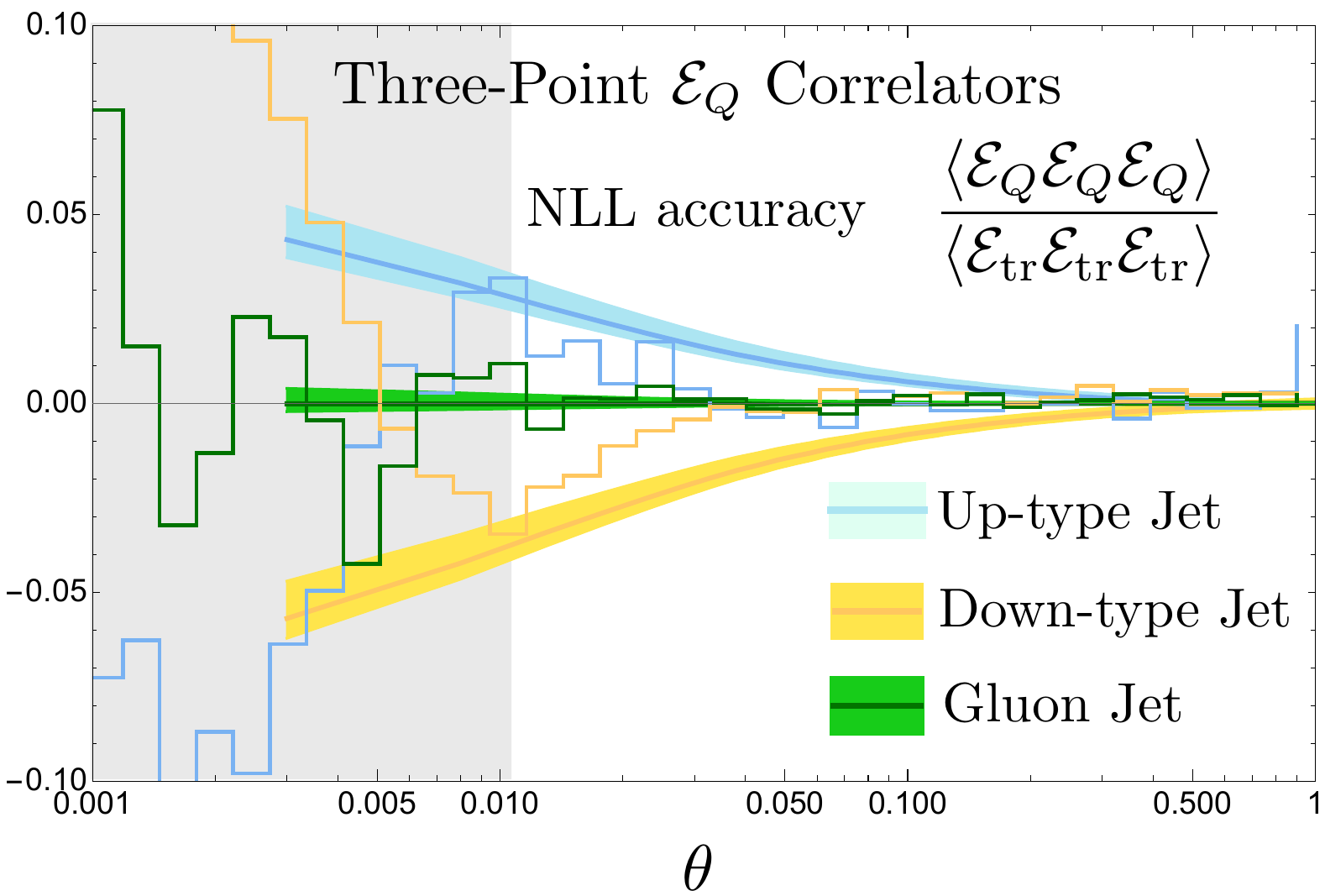} 
\end{center}
\vspace{-0.25cm}
\caption{The scaling of the projected three-point correlator $\langle \cE_\cQ \cE_\cQ \cE_\cQ \rangle$. This observable is $C$-odd. It therefore vanishes on gluon initiated jets, and gives nearly opposite results on up-type and down-type jets. Results of our theoretical calculations are shown in solid, and results from Pythia are shown in the histogram. The transition to the non-perturbative regime is shown in gray. }
\label{fig:EQ_threepoint}
\end{figure}

\begin{figure}
\begin{center}
    \subfloat[]{
\includegraphics[scale=0.29]{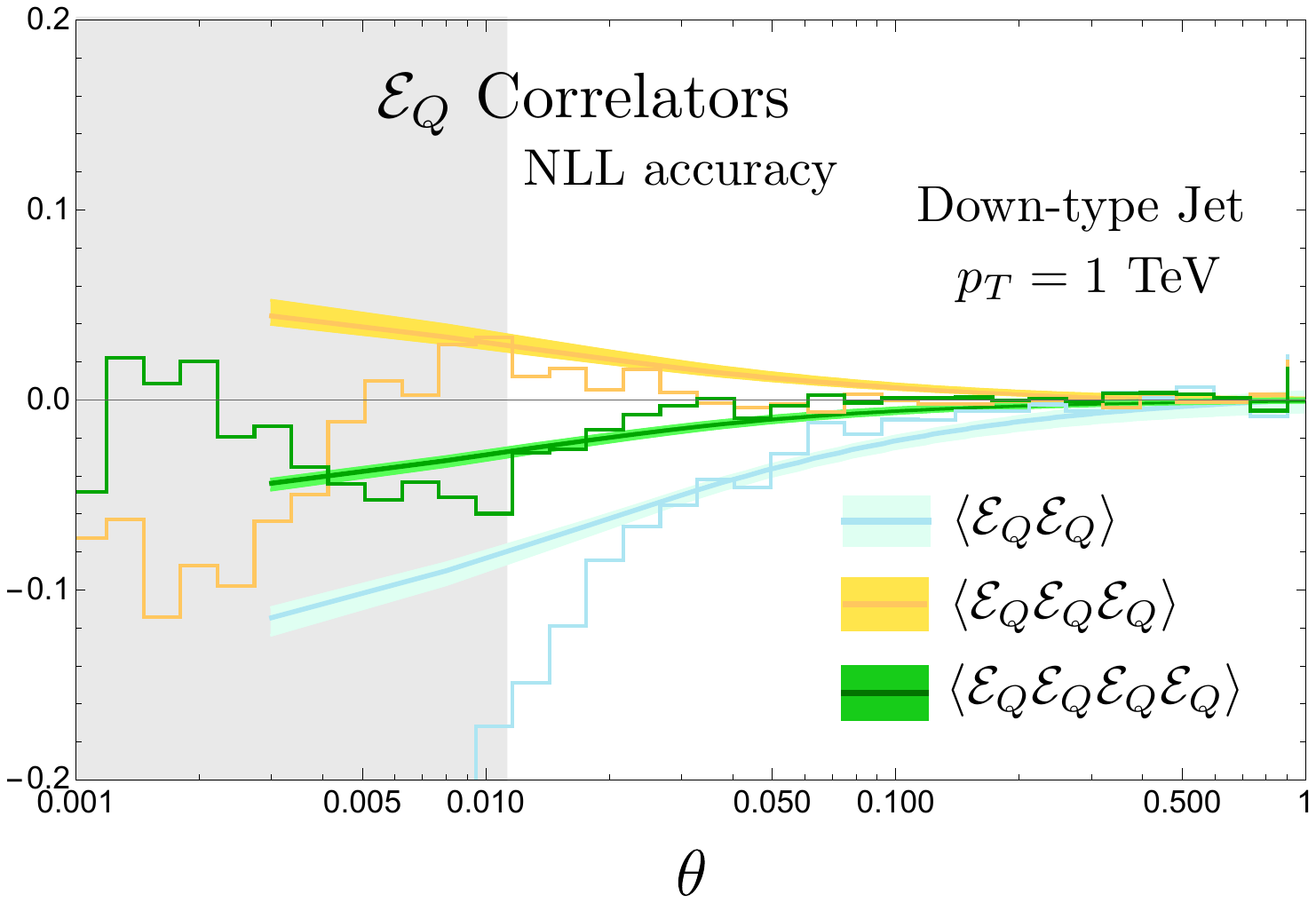} 
  \label{fig:EQ_multipoint_a}
  }\\
      \subfloat[]{
\includegraphics[scale=0.29]{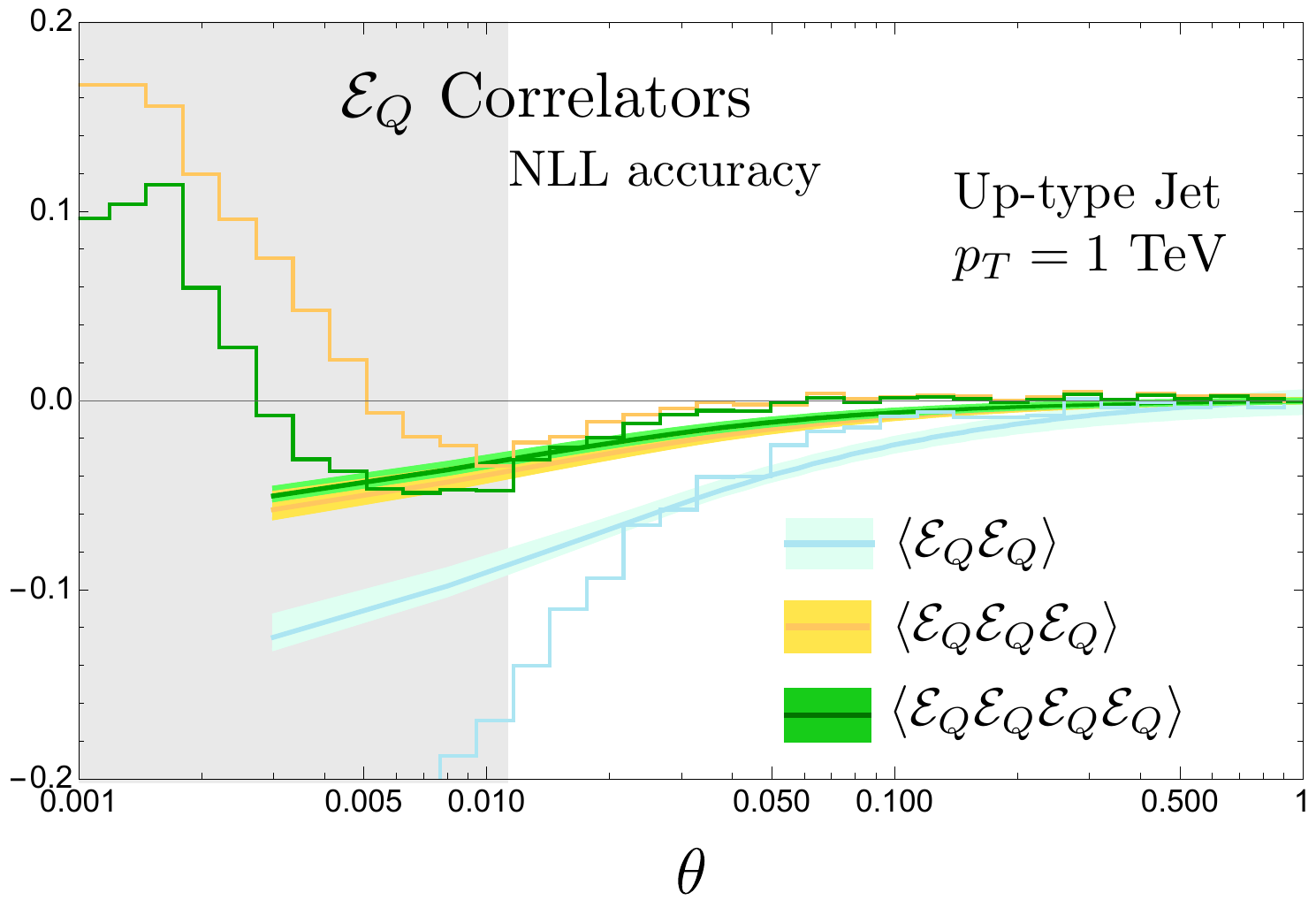} 
  \label{fig:EQ_multipoint_b}
  }\\
        \subfloat[]{
\includegraphics[scale=0.29]{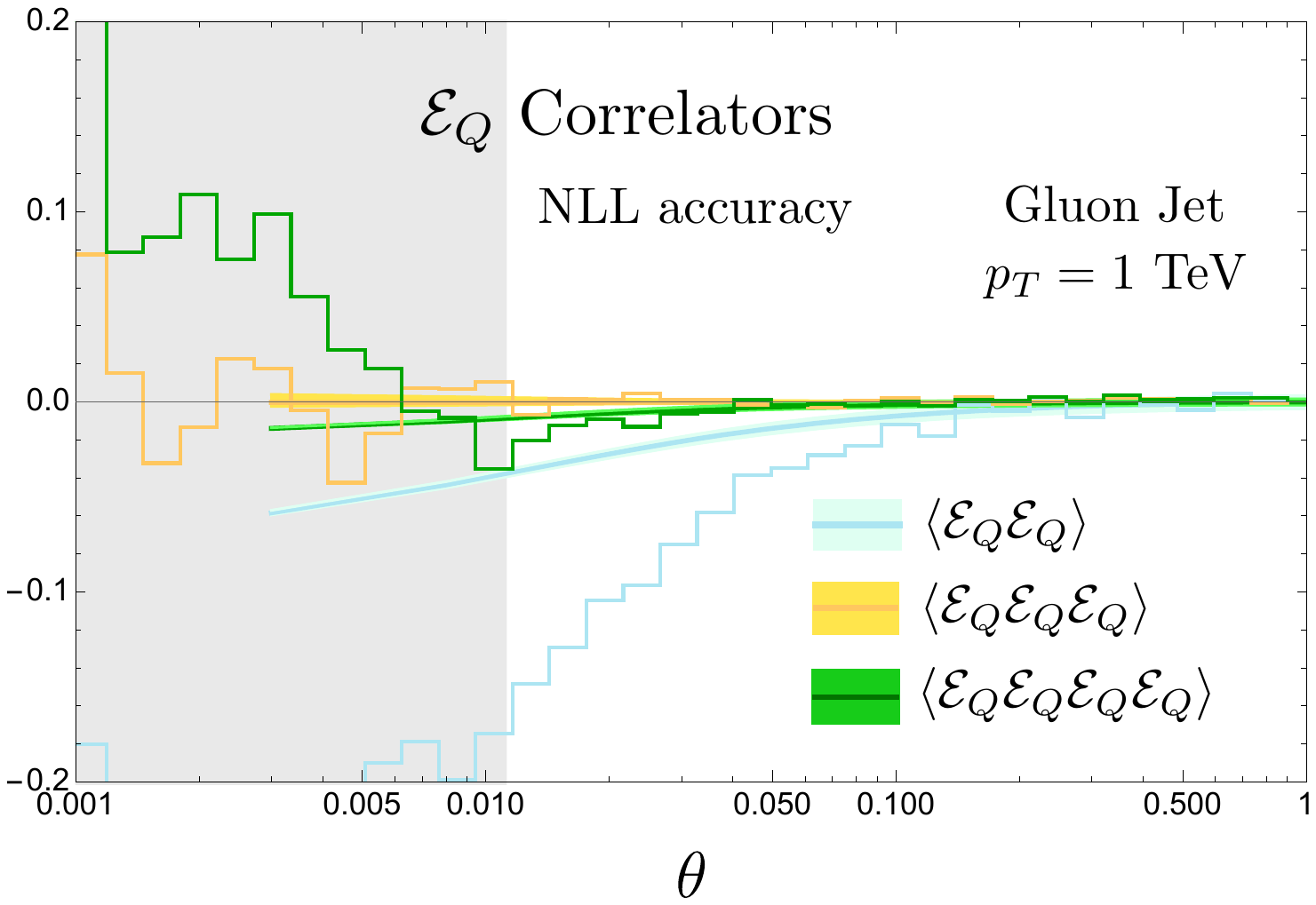} 
  \label{fig:EQ_multipoint_b}
  }
  \end{center}
  \vspace{-0.25cm}
\caption{Multi-point correlators of the $\cE_\cQ$ operator for up-type quarks, down-type quarks, and gluons. These observables are $C$-odd for an odd number of detectors, and $C$-even for an even number of detetors, leading to a qualitatively different behavior on jets initiated by partons of different charges. Results of our theoretical calculations are shown in solid, and results from Pythia are shown in the histogram. The transition to the non-perturbative regime is shown in gray.}
\label{fig:EQ_multipoint}
\end{figure}

We now move onto three-point correlators involving electromagnetic charge. In \Fig{fig:pm_higherpoint} we show different projected three-point energy correlators formed from $\cE_+$ and $\cE_-$ detectors for down type quark jets, up type quark jets, and gluon jets. Slightly different behaviors are observed for the two different combinations of detectors, although the scaling is weak. Perhaps more interesting are higher point correlators of the $\cE_\cQ$ detector. An interesting observable is the $\langle \cE_\cQ \cE_\cQ \cE_\cQ \rangle$ correlator, which is $C$-odd. Because of this it will only give non-vanishing results when computed on electromagnetically charged states. In \Fig{fig:EQ_threepoint} we show the results for up-type quarks, down type quarks, and gluons. We see that the correlator vanishes for gluons, as expected, and gives nearly opposite distributions for up-type and down-type quarks. In \Fig{fig:EQ_multipoint} we show the two-point, three-point and four-point correlators of the $\cE_\cQ$ detector. These exhibit a fascinating behavior due to the fact that the $C$-parity of the observable alternates depending on the number of $\cE_\cQ$ detectors involved in the correlator.  The two-point and four-point correlators are $C$-even, meaning they can be non-zero for gluons, and will give similar results for up and down type jets. On the other hand, the three-point correlator flips between up type and down type quarks and give zero on gluons.

We find it interesting to compare the philosophy of studying correlators of the $\cE_\cQ$ operator, with the traditional approach of studying the total charge of a jet \cite{Field:1977fa}, pursued in \cite{Krohn:2012fg,Waalewijn:2012sv}. The main realization of the program to reformulate jet substructure in terms of correlation functions, as opposed to jet shapes, is that by constraining all the radiation in jets, jet shapes are in fact much more complicated theoretically. Jet shapes are no longer correlators, but rather infinite sums of correlators \cite{Chen:2020vvp}, complicating the connection with the underlying theory. The simpler observables with a more direct connection to the underlying field theory are correlation functions on the energy flow within jets. If we apply these lessons to studying charge, we conclude that instead of studying the total charge within a jet, we should study \emph{correlation functions} of $\cE_\cQ$ within jets. We hope that this reformulation of how to study charge within jets can lead to a similar level of advancement as has been achieved with the transition from jet shapes to energy correlators.

%%%%%%%%%%%%%%%%%%%%%%
\section{Conclusions}\label{sec:conc}
%%%%%%%%%%%%%%%%%%%%%%  

In this paper we introduced the field theoretic formalism necessary to calculate correlation functions $\langle \cE_{R_1}(n_1) \cdots \cE_{R_k}(n_k) \rangle$, of energy flux on restricted sets of quantum numbers, $R_i$. The previously studied track function formalism \cite{Chang:2013rca,Chang:2013iba} allows the calculation of multi-point correlation functions on a particular restricted energy flux  $\langle \cE_{R}(n_1) \cdots \cE_{R}(n_k) \rangle$, but not correlations between energy flux on different subsets of particles. To enable the calculation of these observables, we have introduced a new universal non-perturbative function describing the fragmentation process, the joint track function, and studied its renormalization group evolution and basic properties.  We also showed how its moments appear naturally in a variety of jet substructure observables. 

The main recent advance in the study of jet substructure which makes the introduction of the joint track functions particularly interesting, is an understanding of how to formulate observables that are only sensitive to particular moments of non-perturbative matrix elements.  While track functions have existed for nearly 10 years, it has just been in the last year that they have become practical. This is due to the advent of the energy correlators as a jet substructure observable \cite{Chen:2020vvp}. Factorization theorems for the energy correlators involve only moments of the track functions, which are \emph{numbers}, enabling higher order perturbative calculations on tracks \cite{Li:2021zcf}. In this paper we have shown that this can be greatly extended, and generic correlations characterizing correlations between energy flux carried by hadrons with different quantum numbers, $\langle \cE_{R_1}(n_1) \cdots \cE_{R_k}(n_k) \rangle$, can be described in terms of moments of the joint track functions. This further extends the class of jet substructure and event shape observables to which modern perturbative techniques can be applied.

A key application of the  new non-perturbative matrix elements we have introduced in this paper is in the calculation of energy correlators involving both positive and negative electromagnetic charges, such as $\langle \cE_+(n_1) \cE_-(n_2) \rangle$, or correlators of energy flux weighted by charge $\cE_\cQ$. We presented numerical results for a variety of such correlators, computed at collinear NLL accuracy. These exhibit a variety of fascinating scaling behavior, providing insight into the structure of correlation of charges within jets for the first time. This illustrates the power of combining universal non-perturbative matrix elements with perturbation theory and the renormalization group.  

It will be of particular interest to measure these more general correlators experimentally. We believe that this should be possible, due to the recent measurements of the standard energy correlators \cite{talk1,talk2,talk3}. Indeed, we believe that these correlator based observables may offer several experimental advantages as compared with the proposal of \cite{Chien:2021yol}, which are based on correlations between leading and next-to-leading hadrons. Such an observable is difficult to calculate theoretically, but may also be challenging experimentally, since hadron decays can significantly re-arrange the energy ordering of particles.

We hope that the greatly expanded set of jet substructure observables introduced in this paper will provide new ways of extracting physics from jet substructure at hadron colliders, ultimately leading to new insights into the dynamics of the Standard Model and the strong nuclear force.

%%%%%%%%%%%%%%%%%%%%%%%%%%%%%%%%%%%%%%%%%%%%
\acknowledgments
%%%%%%%%%%%%%%%%%%%%%%%%%%%%%%%%%%%%%%%%%%%%
We thank Helen Caines, Gregory Korchemsky, Andrew Tamis and Youqi Song for useful and motivating discussions. K.L.~was supported by the U.S.~DOE under contract number DE-SC0011090.
I.M.~is supported by start up funds from Yale University.

\bibliographystyle{JHEP}
\bibliography{EEC_ref.bib}

\end{document}